\documentclass[lettersize,journal]{IEEEtran}
\usepackage{amsmath,amsfonts}
\usepackage{algorithmic}
\usepackage{algorithm}
\usepackage{array}
\usepackage[caption=false,font=normalsize,labelfont=sf,textfont=sf]{subfig}
\usepackage{textcomp}
\usepackage{stfloats}
\usepackage{url}
\usepackage{verbatim}
\usepackage{graphicx}
\usepackage{cite}
\usepackage{amssymb}
\usepackage{booktabs}
\usepackage{bm}
\usepackage{color}
\usepackage{verbatim}
\usepackage[table,xcdraw]{xcolor}
\usepackage{array}
\usepackage[switch]{lineno}
 \usepackage{array}
\begin{document}

\title{RDFNet: Regional Dynamic FISTA-Net for Spectral \\ Snapshot Compressive Imaging}

\author{Shiyun~Zhou,
	Tingfa~Xu$^{\dagger}$,
	Shaocong~Dong,
	Jianan~Li$^{\dagger}$
% <-this % stops a space
\thanks{Shiyun Zhou, Tingfa Xu, Shaocong Dong, Xiangmin Li and Jianan Li are with School of Optics and Photonics, Image Engineering \& Video Technology Lab, Beijing Institute of Technology, Beijing 100081, China and the Key Laboratory of Photoelectronic Imaging Technology and System, Ministry of Education of China, Beijing, 100081, China. email: zhoushiyun@bit.edu.cn (Shiyun Zhou), ciom\_xtf1@bit.edu.cn (Tingfa Xu), shaocong@bit.edu.cn(Shaocong Dong), li\_xiangmin@bit.edu.cn(Xiangmin Li), and lijianan@bit.edu.cn (Jianan Li).}% <-this % stops a space
\thanks{Tingfa Xu is with Big Data and Artificial Intelligence Laboratory, Beijing Institute of Technology Chongqing Innovation Center (BITCQIC), Chongqing, 401151, China. e-mail: ciom\_xtf1@bit.edu.cn (Tingfa Xu).}% <-this % stops a space
%\thanks{Jianan Li and Tingfa Xu are the corresponding authors. e-mail: lijianan@bit.edu.cn (Jianan Li), ciom\_xtf1@bit.edu.cn (Tingfa Xu)}}
\thanks{$^{\dagger}$ Correspondence to: Jianan Li and Tingfa Xu. e-mail: lijianan@bit.edu.cn (Jianan Li), ciom\_xtf1@bit.edu.cn (Tingfa Xu)}}

\markboth{Journal of \LaTeX\ Class Files,~Vol.~14, No.~8, August~2021}%
{Shell \MakeLowercase{\textit{et al.}}: A Sample Article Using IEEEtran.cls for IEEE Journals}


\maketitle

\begin{abstract}
    Deep convolutional neural networks have recently shown promising results in compressive spectral reconstruction. Previous methods, however, usually adopt a single mapping function for sparse representation. Considering that different regions have distinct characteristics, it is desirable to apply various mapping functions to adjust different regions' transformations dynamically. With this in mind, we first introduce a regional dynamic way of using Fast Iterative Shrinkage-Thresholding Algorithm (FISTA) to exploit regional characteristics and derive dynamic sparse representations. Then, we propose to unfold the process into a hierarchical dynamic deep network, dubbed RDFNet. The network comprises multiple regional dynamic blocks and corresponding pixel-wise adaptive soft-thresholding modules, respectively in charge of region-based dynamic mapping and pixel-wise soft-thresholding selection. The regional dynamic block guides the network to adjust the transformation domain for different regions. Equipped with the adaptive soft-thresholding, our proposed regional dynamic architecture can also learn appropriate shrinkage scale in a pixel-wise manner. 
    Extensive experiments on both simulated and real data demonstrate that our method outperforms prior state-of-the-arts. 
    %Code will be released soon. 
    Our code and data are available at \url{https://github.com/SherryZhou97/RDFNet}.
\end{abstract}

%With this in mind, we first introduce a regional dynamic {\color[HTML]{00009B}{Fast Iterative Shrinkage-Thresholding Algorithm(FISTA)}} that exploits regional characteristics to derive a dynamic sparse representation.

\begin{IEEEkeywords}
Computational spectral imaging, Compressive hyperspectral reconstruction, Dynamic neural networks, Soft-threshold.
\end{IEEEkeywords}

\section{Introduction}
\IEEEPARstart{H}{yperspectral} image contains large amount of spatial information across a multitude of wavelengths, which makes it enjoy the great potential of wide applications, such as remote sensing~\cite{remote}, medical diagnosis~\cite{medical}, biomedical engineering~\cite{biomedical}, archaeology and art conservation~\cite{archaeology}, food inspection~\cite{food} and environmental monitoring~\cite{environment}.

However, capturing hyperspectral images poses a great challenge since each wavelength needs to be captured separately, which is time consuming and limits the practicality of this technique. Traditional methods of spectral imaging include whiskbroom scanning~\cite{spectrometer1}, pushbroom scanning~\cite{spectrometer2}, and wavelength scanning~\cite{spectrometer3}. Such scanning methods suffer a long spectral image acquisition process, making them inapplicable for large scenes or dynamic recording. To mitigate this, researchers start to explore snapshot spectral imaging~\cite{12}. Early endeavors include integral field spectrometry, multispectral beam splitting, and image-replicating imaging spectrometer~\cite{spectroscopy}. These methods, though achieve multispectral imaging through splitting light~\cite{light}\cite{survey}, still fail to obtain massive spectral channels and require bulky optical systems.

\begin{figure}[tbp]
  \centering
   \includegraphics[width=1.0\linewidth]{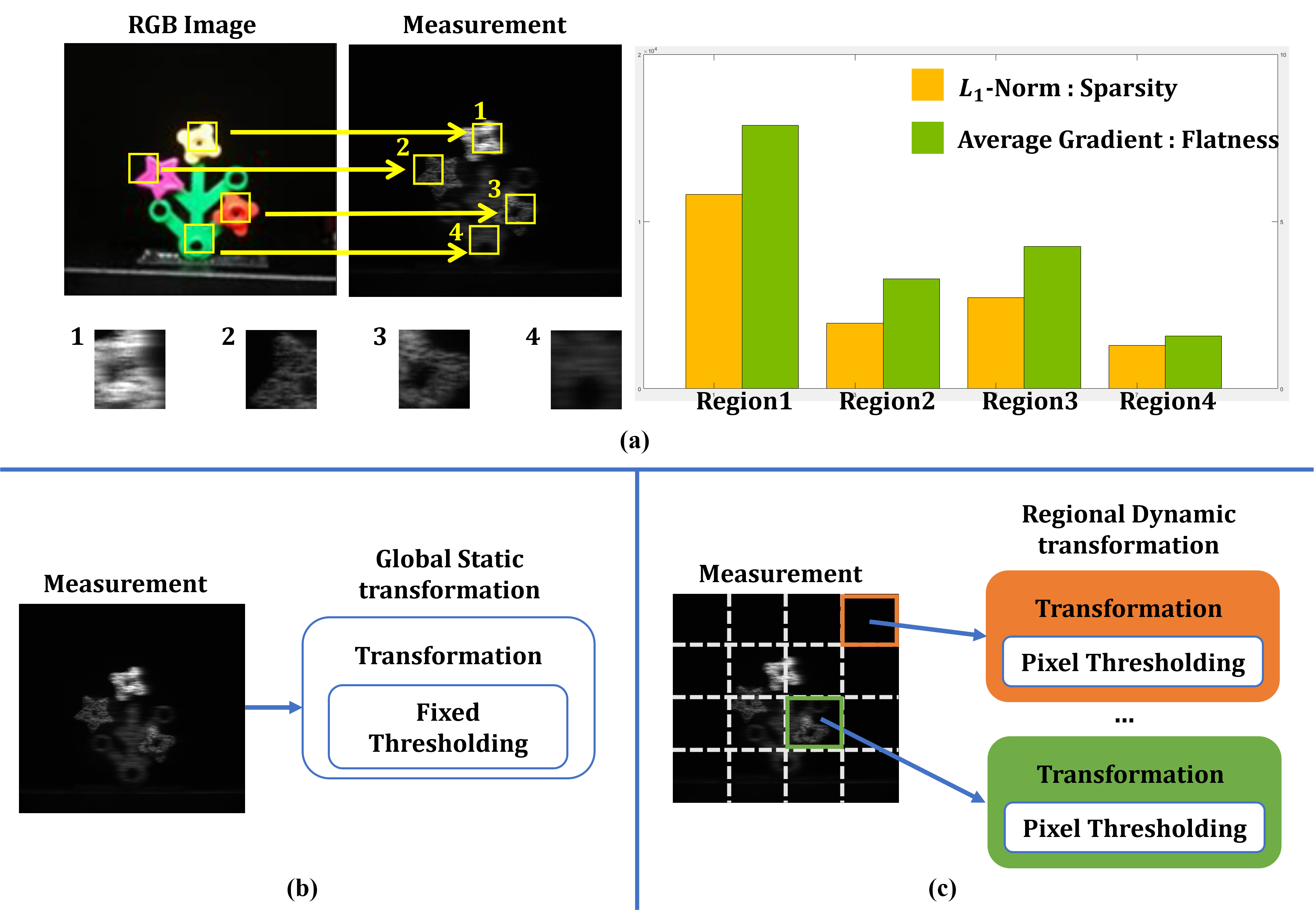}
   \caption{Motivation of this work. (a) 
   The characteristics, \textsl{i.e.}, sparsity and flatness, vary significantly across four regions randomly selected from a real hyperspectral measurement.
   (b) Conventional global static transformation. (c) Our proposed regional dynamic transformation.}
   \label{fig:motivation}
\end{figure}

To tackle the above problems, snapshot compressive imaging (SCI) equipped with advanced compressive sensing (CS)~\cite{CS, CS1} algorithms has received growing attention due to its elegant combination of optics, mathematics, and optimization theory~\cite{light}. Among typical SCI systems, the passive modulation coded aperture snapshot spectral imaging (CASSI) system, which uses a single disperser coded aperture compressive spectral image~\cite{CASSI, SD-CASSI}, stands out due to its low power consumption. It uses a coded aperture to block or filter the input light field, which serves as the encoding process in compressive sensing pipeline~\cite{light}. This process plays a role in information compression, which is flexible in design and provides the prior knowledge for subsequent reconstruction. Different from hardware based encoding, its decoding process largely relies on the computation via designed algorithms. Hence the core challenge of CASSI is to efficiently reconstruct the underlying 3D spectral image from under-sampled 2D measurement.

Traditional reconstruction methods are iterative~\cite{GAP-TV}\cite{adam}\cite{TwIST} and require the designed measurement of the encoding process and other prior knowledge for reconstruction. As a result, the decoding process is computationally expensive and takes minutes or even hours for spectral reconstruction. Moreover, the degradation issue when using limited measurements also hinders the application under resource constrained conditions. To recover the spectra modeled in the complex diffraction process, the powerful deep learning technique is required.

With the rise of Deep Neural Networks (DNN), many studies have attempted to combine DNN with traditional optimization process to replace iterative optimization~\cite{PNP-CASSI, TSA-Net, DNU, lambda-net}. 
Pioneering works~\cite{FISTA-Net, ADMM-Net} tackle this problem by learning a static sparse transformation for the entire image and by using a fixed threshold to obtain the closed-form solution. Nevertheless, we found that different regions of an image have dramatically distinct characteristics. As illustrated in Fig.~\ref{fig:motivation}, the sparsity (measured by $l_1$-norm) and the flatness (measured by average gradient) vary significantly across different regions in real hyperspectral measurement. Inspired by this, we argue that regarding the entire image as a whole and using a single global mapping function may limit the representation of sparse transformation. Different regions need to be transformed into varying sparse domains using different mapping functions based on their unique regional characteristics. 
In addition, the soft-thresholding is used to effectively shrinkage and eliminate the noise-related features in a sparse transformation domain. Similarly, we can dynamically determine the shrinkage scale depending on regions' features. There is much redundancy between high-dimensional information and simple signals in conventional FISTA, a fixed threshold may also limit the denoising capability of the transformation network.

In light of above, this work gives a novel region-based dynamic FISTA~\cite{FISTA} algorithm that uses a regional feature guided weighting approach to dynamically derive the solution in sparse transformation. Guided by the algorithm, we further present a newly designed hierarchical dynamic architecture, dubbed RDFNet, that adopts dynamical multiple mapping functions and uses an efficient and effective strategy to dynamically select the appropriate soft-thresholding of transformation. 

Specifically, RDFNet uses multiple transformation blocks implemented by multilayer perception (MLP) to learn distinct sparse representations. Each of the blocks strictly corresponds to one specific sparse domain. 
Instead of using a fixed threshold, we design a new adaptive soft-thresholding module to automatically determine the threshold, such that the proposed dynamic FISTA transformation block is capable of learning a more appropriate shrinkage scale in each sparse domain. 
Then, we utilize a regional dynamic sub-network to extract the regions' characteristics and generate transform domain weights for each block. 
After that, RDFNet constructs its sparse representation by dynamically assembling multiple fundamental FISTA transformations with regional feature-guided scoring weights. Hence sparse representations are aggregated dynamically for each region.
As a result, our regional dynamic mechanism can greatly enhance the transformation capability of the reconstruction model.
%\JN{add 1-2 sentences to claim the advantage of our design. }

%Extensive experiments on both simulated and real data well demonstrate the superiority of our methods. Compared with the eight existing mainstream methods on 10 scenes of simulation dataset KAIST, our approach achieves state-of-the-art performance by 33.34 dB in average PSNR and 0.956 in average SSIM. 

Extensive experiments demonstrate that the proposed RDFNet outperforms other reconstruction methods on multiple simulation datasets including KAIST~\cite{KAIST}, CAVE~\cite{CAVE} and ICVL~\cite{ICVL}, and also achieves competitive performance on real datasets. In particular, our RDFNet achieves state-of-the-art performance of $33.34$dB in average PSNR and $0.956$ in average SSIM on $10$ scenes of KAIST~\cite{KAIST}. For the natural image dataset ICVL~\cite{ICVL}, our method achieves an average PSNR of 35.51dB.
It also surpasses the previously best-performing DNU by a large margin of $2.9$dB in average PSNR on ICVL~\cite{ICVL} comprised of natural images.
% compared with existing mainstream methods on 10 scenes of simulation dataset KAIST~\cite{KAIST}, 

% For the natural image dataset ICVL~\cite{ICVL}, our method achieves an average PSNR of 35.51dB (2.9dB higher than the second-best DNU).
Moreover, our RDFNet is lightweight with only $1.29$M parameters and runs at a fast inference speed of $0.11$ second per image. These results clearly demonstrate the superiority of RDFNet over prior state-of-the-arts in terms of both accuracy and efficiency.
% Besides, we also measure the model-size, inference speed and the Floating Point Operations(FLOPs) to make the evaluation more comprehensive and convincing. 
% Experimental results show our proposed method RDFNet achieves only 1.29M parameters and inference time of 0.11 second per image, which demonstrates clear superiority over prior state-of-the-arts in terms of both accuracy and efficiency.}

%Compared with the eight existing mainstream methods, our experimental results on the KAIST dataset show an improvement of 33.34dB in PSNR and 0.956 in SSIM with an average runtime of only XXs.Extensive Experiments on both simulated and real data well demonstrate the superiority of our methods.

To sum up, this work makes the following contributions:
\begin{itemize}
    \item We propose a new regional dynamic FISTA algorithm for coded aperture snapshot spectral imaging and design a novel hierarchical dynamic architecture RDFNet.
    \item We present a learnable pixel-wise adaptive soft-thresholding module to automatically determine the shrinkage scale in each transformation block.
    \item We establish new state-of-the-arts on three popular simulation datasets and a real dataset.
\end{itemize}

\section{Related Work}
\subsection{Dynamic Mechanism}
Our work is related to the recent dynamic mechanism. In particular, Chen et al.~\cite{Dynamic} propose a dynamic convolution that aggregates multiple convolution kernels dynamically based on the input. Brabandere et al.~\cite{filter} present a dynamic filter network to dynamically generate position-specific filters on pixel inputs. CondConv~\cite{condconv} generates convolution kernels by combining several filters through a routing function.
% that outputs the coefficients for filter combination. 
Recently, PAConv~\cite{paconv} develop a position adaptive convolution operator with dynamic kernel assembling for point cloud processing. 
However, the region-based dynamic mechanism has not yet been explored in the field of SCI reconstruction.
Zhang et al.~\cite{R2-1} design a weight for each pixel in an image, use the same transformation to perform super-resolution, and add the weights to obtain a mixed transformation for the entire image. 
In comparison, we split the image into regions instead of pixels and perform distinct domain transformations with pixel-level adaptive thresholding for different regions while retaining neighborhood information.
% In different domain transformations, we \R{apply} pixel-level adaptive thresholds.}

\subsection{Learning based Deep Image Prior(DIP)}
With the rise of neural networks, some algorithms try to use the convolution to obtain the DIP but there is no deep network structure, forming a machine learning algorithm.

Bacca J. et al~\cite{R1-1} proposed a network for spectral reconstruction without training according to the ideas of solving ill-posed problems with low rank. It is mainly achieved by analyzing the low-rankness of images at the first layer of the network. Evaluating the difference between minimized compression measurements and predictions by the use of $l_2$. However, it does not really use a deep neural network in the process of solving, but uses several convolutions to help getting the prior of recon. Van Veen D. et al~\cite{R1-2} also proposed an untrained model, which may belong to a kind of machine learning method. The neural network is only used to learn the weight of the prior information, not the way to really obtain the prior information, and the neural network here is not deep, but only uses the volume product. Inspired by the linear mixture model (LMM) for spectral image, Gelvez T. et al~\cite{R1-3} decomposed the image into a matrix, and uses the neural network to learn the weights and features of each matrix as the depth prior of the image for reconstruction. 

In \cite{HyperNet}, DIP is employed as a refinement process of the trained network for the reconstruction of a single image. The other related work is DeepRED~\cite{Deepred}, where DIP is combined with Regularization by Denoising (RED)~\cite{denoisingred}. And the hyperspectral way of using deepred~\cite{Deepred} is~\cite{DIP-HSI}. In fact, most of the processes have nothing to do with the design of deep neural networks, and does not using the characteristics of adjusting the transformation domain for optimizing reconstruction tasks.

\subsection{Deep Learning-based Algorithms}
Inspired by the prevalence of deep learning in the field of high-level visions, some researchers have attempted to use deep convolution neural networks (CNNs) to learn the inverse process. These deep learning-based algorithms can be divided into three streams: End-to-End (E2E)~\cite{U-Net, lambda-net, GAN, TSA-Net}, Plug-and-Play (PNP)~\cite{PNP-ADMM, PNP-video-SCI, PNP-CASSI}, and deep unfolding~\cite{GAP-Net, ADMM-Net, GSM, DNU, low-rank}. \\

\noindent {\bf{End-to-End(E2E):}} E2E-CNNs first applied for its great migration. Both the U-net~\cite{U-Net} and GAN~\cite{GAN} structure has been used for video SCI. The self attention mechanism has been attempted in TSA-Net~\cite{TSA-Net} for spectral SCI. The $\lambda$-Net~\cite{lambda-net}, where a two stage network was proposed. E2E-CNNs enjoy the advantage of fast inference after training, however, it requires a large amount of training data and excessive training time. In addition, E2E-CNN lacks flexibility as well as interpretability.\\

\noindent {\bf{Plug-and-Play(PNP):}} The PnP based algorithms employ pre-trained deep denoising networks as priors and integrate them into the iterative algorithms. Now, applying the well pre-trained denoising networks, such as the FFDNet~\cite{FFDNet}, with ADMM or GAP into SCI leads to fast, flexible and efficient algorithms. The PNP-ADMM~\cite{PNP-ADMM} and PNP-GAP~\cite{GAP-Net} have recently been developed into flexible deep denoisers. A joint reconstruction and demosaicing framework has recently been proposed in~\cite{PNP-SCI} for video SCI and a deep denoiser in~\cite{PNP-CASSI} has shown competitive performance for spectral SCI. However, the pre-trained networks in PnP methods are fixed without re-training, therefore limiting the performance.\\

\noindent {\bf{Deep Unfolding:}} Deep unfolding merges the advantages of the iterative optimization and E2E-CNNS by training a concatenation of small CNNs to simulate the iterative operations in traditional optimization, where each phase is referred to as a stage. Optimization-based update rules are used to connect these phases and train in an end-to-end fashion. It is somewhat interpretable. Since the small CNNs are independent of the sensing matrix, they can be trained with a smaller dimension than the size of the desired signal, which makes them both training and testing faster than E2E-CNN.

Most recently, the GAP-net proposed in~\cite{GAP-Net} has achieved good results in both video and spectral SCI. A deep unfolding based on the Gaussian scale mixture model has been developed in~\cite{GSM} for spectral SCI reconstruction. DNU~\cite{DNU} has contributed to the introduction of a new prior for optimization. Zhang and Wang~\cite{low-rank} first learned the tensor low-rank prior of hyperspectral images in the feature domain by DNN to promote the reconstruction quality.

Nonetheless, these methods still show limitations in modeling sparsity representations. Besides, the guidance of regional characteristics for adjusting the reconstruction transformation domain is under-studied.

% \R{Nonetheless, these methods show limitations in modeling  spatial-spectral correlations. Besides, the guidance of adjusting the recon transformation domain based on the regional characteristics is under-studied.}

\section{Method}
We first revisit the typical CASSI observation model, then introduce our regional dynamic FISTA algorithm, and finally elaborate our regional dynamic FISTA network (RDFNet).

\begin{figure}[t]
  \centering
  \includegraphics[width=1.0\linewidth]{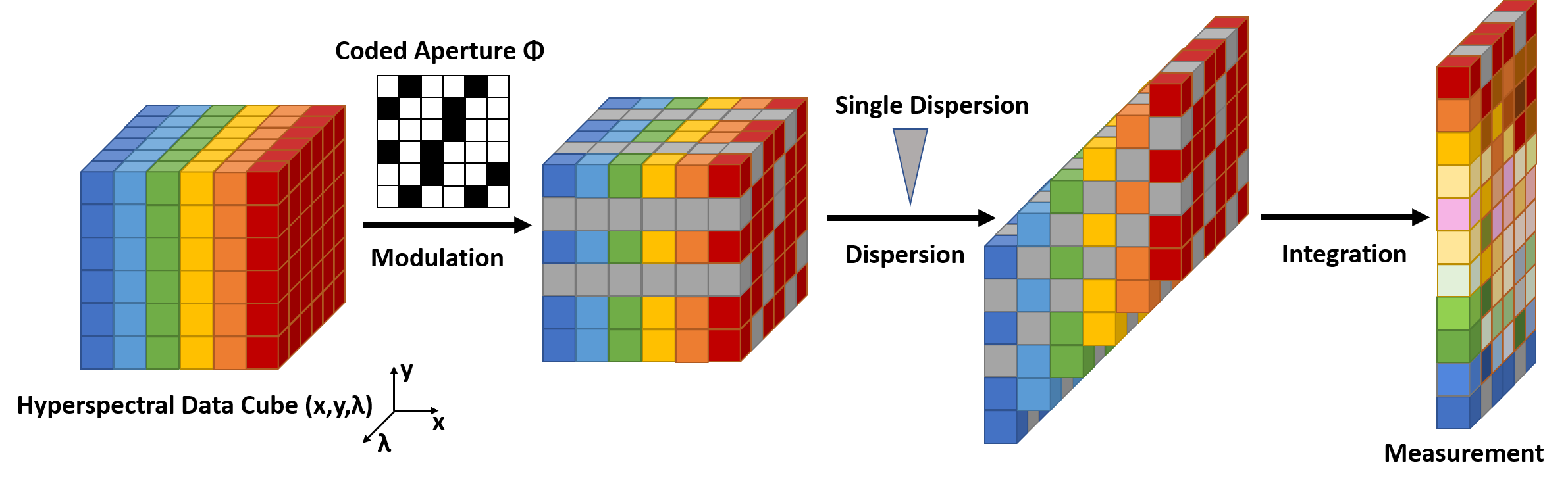}
   \caption{Schematic diagrams of spectral snapshot compressive imaging, a.k.a., coded aperture snapshot spectral imaging (CASSI) system. The spatial-spectral datacube is first modulated by a fixed physical mask and then the modulated datacube is sheared by a disperser. The 2D coded measurement thus includes the information of the spectral datacube, which is the desired 3D signal.}
   \label{CASSI}
\end{figure}

\subsection{CASSI Observation Model}
Spectral snapshot compressive imaging (SCI) sysmtem comprises of a hardware encoder and a software decoder. The encoder denotes the optical system that compresses 3D data cube $(x,y,\lambda)$ to a snapshot measurement on a 2D detector. The decoder denotes the reconstruction algorithm used to recover the 3D data cube from the snapshot measurement. 
Here, we focus on the coded aperture snapshot spectral imaging (CASSI) system that uses a fixed mask and a disperser to implement band-wise modulation.

As shown in Fig.2, each spatial position of the scene is modulated by a coded aperture (mask) that blocks or unblocks the incoming light. Then the coded spectral scene passes through the prism to introduce a horizontal shifting. Finally, the coded shifted spectral scene is integrated along the spectral axis by the detector, resulting in 2D compressed measurement.

Following the theory in \cite{SCI}, let $X\in \mathbb{R}^{N_x\times N_y \times N_{\lambda}}$ denote the 3D spatial-spectral cube and $M^0 \in (0,1)^{N_x\times N_y}$ the physical mask used for signal modulation. We use $X'\in \mathbb{R}^{N_x\times N_y \times N_{\lambda}}$ to represent the modulated signal where images at different wavelengths are modulated separately by the same mask. For $n_\lambda \in \left\{1, ...., N_\lambda\right\}$, we have:
\begin{equation}
    X'(:,:,n_\lambda) = X(:,:,n_\lambda)\odot M^0,
\end{equation}
where $\odot$ represents element-wise multiplication.

Next comes the disperser, which disperses the light to different spatial locations based on their wavelengths. After the modulated cube passes the disperser, $X'$ is tilted and considered to be sheared along the $y-$axis.
We use $X''\in \mathbb{R}^{N_x\times(N_y+N_\lambda-1)\times N_\lambda}$ to denote the tilted cube and assume $\lambda_c$ to be the reference wavelength. That is, image $X'(:,:,n_{\lambda_c})$ is not sheared along the $y-$axis, we hence have:
\begin{equation}
 X''(u,v,n_{\lambda}) = X'(x, y + d(\lambda_n - \lambda_c), n_\lambda),   
\end{equation}
where $(u,v)$ indicates the coordinate system on the detector plane, and $\lambda_n$ is the wavelength of channel $n_\lambda$. 
Here, $d(\lambda_n - \lambda_c)$ signifies the spatial shifting for channel $n_\lambda$. The compressed measurement at the detector $y(u,v)$ can thus be modeled as
\begin{equation}
    y(u,v) = \int_{\lambda_min}^{\lambda_max}x''(u,v,n_\lambda)d\lambda,
\end{equation}
since the sensor integrates all the light in the wavelength range $[\lambda_{min}, \lambda_{max}]$, where $f''$ is the continuous representation of $F''$. In discretized form, the captured 2D measurement $Y \in \mathbb{R}^{N_x\times(N_y+N_\lambda-1)}$ is
\begin{equation}
    Y = \sum_{n_\lambda=1}^{N_\lambda}X''(:,:,n_\lambda) + \epsilon,
\end{equation}
which is a compressed frame containing information of all the modulated spectral channels and $\epsilon \in \mathbb{R}^{N_x\times(N_y+N_\lambda-1)}$ represents the measurement noise.
For simplicity purpose, we denote $M \in R^{N_x\times(N_y+N_\lambda-1)\times N_\lambda}$ as the shifted version of the physical mask corresponding to different wavelengths,
\begin{equation}
    M(u,v,n_\lambda) = M^0(x, y + d(\lambda_n - \lambda_c)).
\end{equation}
Similarly, for each signal frame at different wavelengths, the shifted version $\tilde{X} \in R^{N_x\times(N_y+N_\lambda-1)\times N_\lambda}$ is
\begin{equation}
    \tilde{X}(u,v,n_\lambda) = X(x, y + d(\lambda_n - \lambda_c), n_\lambda).
    \label{eq6}
\end{equation}
Based on the above, measurement $Y$ can be represented as
\begin{equation}
   Y = \sum_{n_\lambda=1}^{N_\lambda}\tilde{X}(:,:,n_\lambda)\odot M(:,:,n_\lambda) + \epsilon.
\end{equation}
This corresponds to the encoding process of SCI in Fig.2. Note that the 3D mask $M$ can be obtained by calibration. Given the solved $\tilde{X}$, we can obtain the desired 3D cube by shifting it back to $F$ based on the relationship in Eq.(\ref{eq6}),
\begin{equation}
    x = [x_1^T, ... ,x_{N_t}^T]^T,
    \Phi = [D_1, ..., D_{N_t}]^T,
\end{equation}
where $x_k = vec(X_k)$ represents the vectorization of frame $k$ and $D_k = Diag(vec(M_k))$ is a diagonal matrix with diagonal elements vectorized of $M_k$. We obtain the forward model
\begin{equation}
    \bm{y} = \bm{\Phi x} + \epsilon,
\end{equation}
which is the core problem of spectral SCI reconstruction.
%Fig.~\ref{CASSI} illustrates a schematic of CASSI. The spectral information is first spatially modulated by a coded aperture with a fixed pattern and then spectrally dispersed by the dispersive prism before being detected by the detector. The core problem of spectral SCI reconstruction is to solve the ill-posed problem as follows:
%\begin{equation}
    %\bm{y} = \bm{\Phi x} + \epsilon,
%\end{equation}
%where $\bm{y}$ is the measurement that the 2D detector received, $\bm{\Phi}$ is the option modulating by the disperser, $\bm{x}$ is the desired reconstruction signal, and $\epsilon$ denotes the noise in reconstruction.
Conventional methods~\cite{sparsity, tv, prior} usually employ a regularization term $R(\bm{x})$ as prior to constrain the solution in desired signal space. These algorithms aim to find an estimated $\bm{\Bar{x}}$ of $\bm{x}$ by solving the following problem:
\begin{equation}
    \bm{\Bar{x}} = \arg\min_{\bm{x}}\frac{1}{2}||\bm{y} - \bm{\Phi x}||_2^2 + \lambda R(\bm{x}),
    \label{eq2}
\end{equation}
where $\lambda$ is a parameter to balance between the fidelity and the regularization term. Eq.(\ref{eq2}) is usually solved by iterative algorithms with various image priors of $R(\bm{x}) $ including sparsity~\cite{sparsity}, total variation~\cite{tv}, deep denoising prior~\cite{PNP-ADMM,PNP-CASSI}, autoencoder prior\cite{U-Net}, \textit{etc}. 

\subsection{Regional Dynamic FISTA Algorithm}
% Based on the CASSI observation model, 
Given the measurement $\bm{y}$ and the modulate mask $\bm{\Phi}$, the problem of reconstructing hyperspectral image $\bm{x}$ can be solved by LASSO optimization~\cite{lasso}. Using $l_1$-norm to impose sparsity constraint for coefficients~\cite{FISTA-Net}, the reconstruction problem in Eq.(\ref{eq2}) can be converted as
\begin{equation}
    \bm{\Bar{x}} = \arg\min_x\frac{1}{2}||\bm{y} - \bm{\Phi x}||_2^2 + \lambda ||\bm{\Psi x}||_1,
    \label{eq3}
\end{equation}
where $\bm{\Psi x}$ denotes the coefficients in the transformation domain. By introducing an auxiliary parameter $\bm{r}^k$, the unconstrained optimization in Eq.(\ref{eq3}) can be solved by iterative steps~\cite{FISTA}:

    \begin{equation}
        \bm{x}_k = \arg\min_x\frac{1}{2}||\bm{x} - \bm{r}^k||_2^2 + \lambda||\bm{\Psi x}||_1,
        \label{iter1}
    \end{equation}
    \begin{equation}
        \bm{r}^k = \bm{z}^k - \rho\bm{\Phi}(\bm{\Phi z}^k - \bm{y}),
    \end{equation}
    \begin{equation}
        t^{k+1} = \frac{1 + \sqrt{1 + 4(t^k)^2}}{2},
    \end{equation}
    \begin{equation}
        \bm{z}^{k+1} = \bm{x}^k + (\frac{t^k-1}{t^k+1})(\bm{x}^k - \bm{x}^{k-1}),
        \label{iterend}
    \end{equation}
where $k\geq1$, $\bm{z}^1 = \bm{x}^0, t^1 = 1$, $\rho$ represents the step size. $\bm{z}^{k+1}$ is a new strating point for next iteration. In each step, we directly utilize the updated $t^k$ to calculate $\bm{z}^k$.

\begin{figure*}[htbp]
  \centering
  \includegraphics[width=0.85\linewidth]{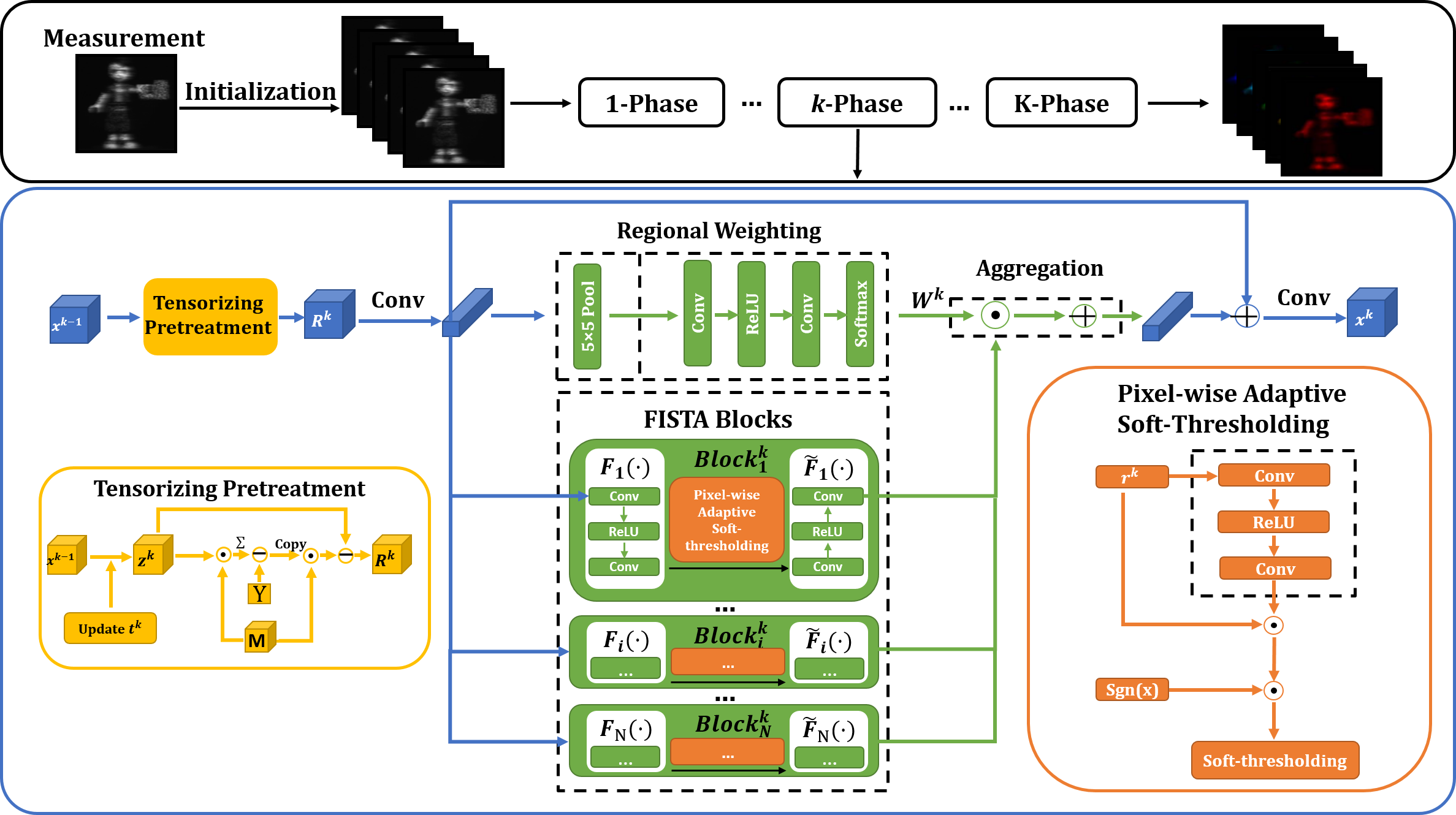}
   \caption{Overall architecture of RDFNet. The upper part demonstrates the data flow in RDFNet, containing K phases. The bottom part is the detailed network implementation of a phase, including pretreatment,  dynamic FISTA blocks equipped with pixel-wise adaptive soft-thresholding module, and a sub-network for regional dynamic aggregation. 
   }
   %including the pretreatment and our proposed dynamic FISTA block and adaptive thresholding.
   \label{Structure}
\end{figure*}
 
%  Conventional FISTA algorithm\cite{FISTA} regards the image as a whole, using a single mapping function to process in a simply global static transformation. While we found it varies significantly between different regions in a whole measurement, we are focus on using multiple mapping functions for various regions to realize different transformations, through the region-level characters to adjust dynamically.
Conventional FISTA algorithm~\cite{FISTA} regards the image as a whole and performs simple global static transformation using a single mapping function. However, we found that there exist significant differences among different regions in a measurement. Hence we are dedicated to applying distinct mapping functions for different regions based on their unique characteristics to realize region-adaptive transformation.

Based on the tensorlization operations~\cite{FISTA-Net},  we re-reference the theoretical process of RDFNet to match our regional dynamic transformation and pixel-wise soft-thresholding.
Specifically, we divide input $\bm{r}^k$ into a series of regions $\bm{r}_i^k \in \left\{\bm{r}_1^k, ..., \bm{r}_M^k\right\}$, and process each individual region using a dynamic mapping function determined by the region's characteristics. By using $F(\cdot)$ to learn the sparsest representation of spectral images, we can obtain the regional results $\bm{x}_i^k \in \left\{\bm{x}_1^k, ..., \bm{x}_M^k\right\}$ based on the relationship in Eq.(\ref{eq3}):
\begin{equation}
        \bm{x}_i^k = \arg \min_{\bm{x}_i^k} \frac{1}{2}||\bm{x}_i^k - \bm{r}_i^k||_2^2 + \lambda||F(\bm{x}_i^k)||_1.
    \label{function_psi}
\end{equation}
Following the Parseval Theorem
\begin{equation}
    ||\bm{Da} - \bm{Db}||_2 = ||\bm{a} - \bm{b}||_2,
\end{equation}
where $\bm{D}$ is an orthonormal transformation matrix. Eq.(\ref{function_psi}) can be converted as
\begin{equation}
        \bm{x}_i^k = \arg \min_{\bm{x}_i^k} \frac{1}{2}||F(\bm{x}_i^k) - F(\bm{r}_i^k)||_2^2 + \lambda||F(\bm{x}_i^k)||_1.
    \label{function_F}
\end{equation}
%use multiple mapping functions to learn different transformations, dynamically adjusting to a mixture transformation.
%  Once obtain the functional form of the inverse problem, the FISTA algorithm\cite{FISTA} unfolding the number of iterations in K phases and then generally used a mapping function to emulate the iterative operations in a phase. 
%Once FISTA\cite{FISTA} obtains the functional form of the inverse problem, it unfolds the number of iterations of K phases and then generally uses the mapping function to emulate an iterative operation of a phase. 
According to the soft-thresholding theory~\cite{soft-threshold}, we adopt soft-thresholding operator to obtain the closed-form solution for each region:
\begin{equation}
    \bm{x}_i^k = \hat F'(soft(\hat F(\bm{r}_i^k), \lambda)),
\end{equation}
where $\hat F', \hat F$ are the mixture dynamic transformation. By summing up the regional results $\bm{x}_i^k \in \left\{\bm{x}_1^k, ..., \bm{x}_M^k\right\}$, we can obtain the final solution $\bm{x}^k$.

To achieve the adjustment of regional dynamical transformation $(\hat F(\cdot),\hat F'(\cdot))$, we first design multiple mapping functions $\left\{(F_i(\cdot), F'_i(\cdot))\right\}_{i=1}^N$  to represent different fundamental transformations. Then we derive several regional characteristic-driven weights $\bm{w}_i^k$ corresponding to each mapping function. Hence the transformation can be dynamically adjusted according to the weights. The solution can be calculated as:

\begin{equation}
    \bm{x}^k = \sum_{i=1}^N\hat F_i'(soft(\hat F_i(\bm{r}_i^k), \lambda))\cdot \bm{w}_i^k.
\end{equation}

For soft-thresholding, we aim to learn an adaptive threshold for each pixel within a region. Specifically, we design a pixel-wise adaptive soft-thresholding $soft^{\bm{\tau}_i^k}$ by using $sgn(\bm{x})$ to shrinkage every signal pointed among transformations:
\begin{equation}
soft^{\bm{\tau}_i^k}(\bm{x}) = sgn(\bm{x})(|\bm{x}| - \bm{\tau}_i^k).\\
\end{equation}
where $\bm{\tau}_i^k$ is the adaptive soft thresholding determined by regions' characteristic $\bm{r}_i^k$. 

%{\color[HTML]{00009B}$R_k$ is the tensor form of auxiliary parameter $r$ in the $k-th$ phase}

Inspired by the skip connection in ResNet~\cite{ResNet}, we obtain the closed-form solution of Eq.(\ref{function_F}) as:
\begin{equation}
    \bm{x}^k = \sum _i^N (F'_i soft^{\tau}_{i^k} (F_i(\bm{r}_i^k),\lambda)\cdot \bm{w}_i^k + \bm{r}_i^k).
    \label{solution}
\end{equation}
Hence, we can achieve the desired solution in a learnable manner.

\subsection{Regional Dynamic FISTA-Net}
Next, we design a novel Regional Dynamic Network (RDFNet) to implement the above regional dynamic FISTA algorithm.

\noindent\textbf{Overview.}
Fig.~\ref{Structure} shows the overall architecture of RDFNet, which performs the following workflow:
\textit{i)} split the measurement into a 3D data cube to initialize $x$;
\textit{ii)} complete the iterative steps of FISTA algorithm in Eq.(\ref{iter1})-Eq.(\ref{iterend}) through tensorizing pretreatment and convert into tensor form;
\textit{iii)} extract regional characteristics to generate the weights for guiding transformations;
\textit{iv)} learn multiple fundamental transformations using hierarchical dynamic blocks with pixel-adaptive soft-thresholding;
\textit{v)} assemble different fundamental transformations using the regional-based weights, aggregating into final output.

Specifically, we first slide a $H\times W$ extraction window on the input 2D measurement of size $H\times (W + L - 1 )$ with slide step of one pixel, and split the input into $L$-channel image of size $H\times W$. Then the split sub-images are fed into the tensorizing pretreatment as stated in~\cite{FISTA-Net} to transfer the iteration from vector to tensor form to  reduce interference time and memory footprint.

% ,which will save much inference time and memory footprint.

Based on the deep-unfolding framework, we propose a novel deep architecture for solving the proximal mapping problem of compressive sensing reconstruction by using a dynamic nonlinear sparsifying transformation at each iterative phase. It contains three main components.

% \R{The proposed regional weighting module can extract the regional characteristics to generate region-wise dynamic weights for guiding transformations by fusing the outputs of multiple dynamic blocks with weighted summation. }

The proposed regional weighting module extracts regional characteristics to generate a region-wise dynamic weight to guide the optimal sparsifying transformation for each region. Different from previous iterative methods that perform a single fixed transformation for the entire image, the developed multiple dynamic blocks aim at learning different fundamental transformations for different regions by exploiting their corresponding unique characteristics. Besides, an adaptive threshold module is designed in each dynamic block to learn pixel-wise adaptive soft-thresholds.
% by fusing the outputs of multiple dynamic blocks with weighted summation.
% by exploiting the regional characteristics.

% Besides, an adaptive threshold module is designed in each of the dynamic blocks for learning a pixel-wise adaptive soft-threshold.

% Besides, an adaptive threshold module is designed for learning a pixel-wise adaptive soft-threshold in each of the dynamic blocks.

% \R{Considering that previous iterative methods solely consider fix transformation via same convolutions, the developed multiple dynamic blocks are designed for learning different fundamental transformation by exploiting the regional characteristics of the compressive measurement.}

% \R{In the end, we merge the different fundamental transformations with the regional-based weights to obtain the final output.}
% \R{In the end, we fuse the outputs of multiple dynamic blocks with the region-based weights to obtain the final output.}

% we fuse the multiple dynamic blocks' outputs with a summation of region-based weights to obtain the final output.

Finally, we merge the outputs of multiple dynamic blocks via a summation of region-based weights to obtain the final output.
% In the end, we merge the outputs of multiple dynamic blocks with a summation of region-based weights to obtain the final output.
% \textcolor{blue}{In the end, we fuse the outputs of multiple dynamic blocks with region-dependent weighted summation to obtain the final output.}

% i) The measurement is split into a 3D data cube, following the FISTA pretreatment to initialize x. ii) We proposed a regional dynamic aggregation mechanism to extract the region character and assemble them with the different transformations, then obtain the results by aggregation them up. iii) We design a set of hierarchical dynamic FISTA blocks to learn the mapping functions of sparse representation. Moreover, we utilize a sub-network to adaptively determine the pixel-wise thresholds for each region in transformation.

%\textcolor{color}{the speed up operations in FISTA algorithm are modular as a first pretreatment in each stage. Then, considering the close spectral correlations among the adjacent channels, we use a convolution layer to extract the information among spectral channels and learn the spectral-side's sparse representation. For simplicity, we use the kernel size $(3\times3)$ in convolution layers and increase the number of channels from 28 to 64.}\JN{the above process acts as an essential step for the whole network, make it an independent subsection.}

\noindent\textbf{Tensorizing Pretreatment.}
RDFNet takes the measurement as input and splits it into a 3D data cube. Then, we use the tensorizing pretreatment module to implement the iteration steps of FISTA~\cite{FISTA-Net} algorithm and convert the data form into tensor.
Inspired by video FISTA-Net\cite{FISTA-Net}, consider $Z^k$, $R^k$, and $X^k$ as the tensor form of $z^k$, $r^k$ and $x^k$, respectively. The tensor form of Eq.(\ref{function_F}) becomes:
\begin{equation}
    \bm{X}_i^k = \arg \min_{\bm{X}_i^k} \frac{1}{2}||F(\bm{X}_i^k) - F(\bm{R}_i^k)||_2^2 + \lambda||F(\bm{X}_i^k)||_1.
\end{equation}
After the iterations Eq.(\ref{iter1})-Eq.(\ref{iterend}) of FISTA~\cite{FISTA-Net}, the solution of $X_k$ is:
\begin{equation}
    \bm{X}^k = \sum _i^N (F'_i soft^{\tau}_{i^k} (F_i(\bm{R}_{i^k}),\lambda)\cdot \bm{w}_{i^k} + \bm{R}_{i^k}).
\end{equation}
Considering the close spectral correlations existing among adjacent channels, we learn a linear embedding to extract the information among spectral channels:
\begin{equation}
    \bm{R}^k = L(\bm{R}^k).
\end{equation}
Here we use a $(3\times3)$ convolution to implement the embedding $L(\cdot)$ which increases the number of channels from $28$ to $64$.

% implemented by a $(3\times3)$ convolution layer

% we use a convolution layer to extract the information among spectral channels and learn the spectral-side sparse representation as in FISTA-Net~\cite{FISTA-Net}

% For simplicity, we use the kernel size of $(3\times3)$ in convolution layers 

\noindent\textbf{Design of Dynamic Block.}
After the above pre-processing, we introduce the body parts of RDFNet, the dynamic block. We use a set of $N$ parallel branches to learn different fundamental transformations. Each branch is equipped with an adaptive soft thresholding, which is suitable for the spatially varying signals contained in hyperspectral images.

Obviously, the number of dynamic blocks $N$ plays an important role. A larger $N$ contributes to more diversified domain transformation for sparse representation. Nevertheless, too many transformation domains may lead to redundancies and cause heavy memory and computational overhead. We find that setting $N=3$ is appropriate, which is discussed in Sec. ~\ref{ablation} b)
\paragraph{Fundamental Transformation}
In each dynamic block, we use multilayer perceptions (MLPs) comprised of two convolutional layers and an activation layer to learn the fundamental transformation denoted by $F(\cdot)$,
\begin{equation}
    F(\bm{X}_i^k) = \bm{\omega_2}(\sigma(\bm{\omega_1}(\bm{X}_i^k) + \bm{b}_1)) + \bm{b}_2.
\end{equation}
Each MLPs strictly corresponds to the transformation function $F(\cdot)$ and the inverse transformation function $F'(\cdot)$. $\sigma(\cdot)$ is implemented by a Rectified Linear Unit (ReLU) activation layer.
Besides, we utilize a symmetry constraint~\cite{FISTA-Net} to ensure the two MLPs' use are inverse in a dynamic block.
\begin{equation}
    \arg \min _ {\bm{X}} ||F'(F(\bm{X})) - \bm{X}||^2_2.
\end{equation}
The inverse transformation function takes $F(\bm{X})$ as input and makes the output as close as possible to $\bm{X}$, thus guaranteeing the two MLPs reciprocal to each other.
% \JN{Taking the output $F(\bm{X})$ as the input of inverse transformation function $F'(\cdot)$. Its difference with the original $\bm{X}$ is kept as small as possible to guarantee the two MLPs reciprocal to each other. }
%\JN{Taking the output $F(\bm{X})$ as the input of inverse transformation function $F'(\cdot)$ measures the results with the initial input $\bm{X}$. Keep it as small as possible to guarantee the inversely of two MLPs. }
%\\= \arg\min_x[\omega_4(ReLU(\omega_3F(x) + b_3)) + b_4]

%If we can reduce to get the input of $F(\cdot)$, then we prove that the two mlps are inversly.

% \begin{equation}
% \left\{
% \begin{array}{lcl}
%     F(x) = \omega_2(ReLU(\omega _1x + b_1))+b_2\\
%     F'(x) = \omega_4(ReLU(\omega_3x + b_3)) + b_4\\
%     F'(F(x)) = x\\
    
% \end{array} \right.
% \end{equation}

\paragraph{Pixel-adaptive Soft-thresholding}
We next adopt soft-thresholding\cite{soft-threshold} to remove noise-related features in the sparse transformation domain.
%Once obtain the ouput of transformation function $F(\cdot)$
%we...
The region-based soft-thresholding used in RDFNet transformation can be expressed as:
\begin{equation}
\bm{\tau}_i^k = T(\bm{R}_i^k).
\end{equation}
As illustrated in Fig.~\ref{Structure}, we design a new specialized sub-network to automatically determine the threshold by exploiting the relationship $T(\cdot)$ between regional input $\bm{R}_i^k$ and the threshold. Specifically, given the output of pretreatment $\bm{R}^k$, we use two convolutional layers and a link activation function $\sigma(\cdot)$ to learn the mapping function of $T(\cdot)$, resulting in the prediction of the scaling parameter for each pixel. Besides, the output of the sub-network is scaled to the range of (0, 1), such that the resulting threshold is positive and kept within a reasonable range to prevent the output features from being all zeros:

%\JN{we use two convolutional layers and a link activation function $\sigma(\cdot)$ to predict the scaling parameter for each pixel as the mapping $T(\cdot)$.}

\begin{equation}
    \bm{\tau}_i^k = \bm{\omega}_2(\sigma(\bm{\omega}_1 \bm{R}_i^k + \bm{b}_1))+\bm{b}_2 \in (0, 1).
\end{equation}
Here we implement the activation function by ReLU.
Consequently, the region characteristics adaptively guide the shrinkage scale of every point in the region.
%In this way, the shrinkage scale of every point in the region is adaptively guided by the region characteristics.

% Soft thresholdinging is used to remove noise-related features. As illustrated in~\cref{Structure}, we design a new specialized sub-network as trained modules to automatically determine the thresholdings. It takes the feature from the after feature fused $r^k$ as input. They are using two convolution layers and a link activation function ReLU to predict the scaling parameter for each pixel. The output of the sub-network is scaled to the range of (0, 1) by using a softmax layer. Thus the thresholdings can be positive and kept in a reasonable range, thereby preventing the output features from being all zeros. 

% In summary, the thresholding used in RDFNet is expressed as follows:
% \begin{equation}
%     \tau^k = \omega_2(ReLU(\omega_1 r^k + b_1))+b_2.
% \end{equation}

% Finally, adopt the adaptive threshold $\bm{\tau} ^k$ with $sgn(\bm{x})$, that's completed the shrinkage process of the signals, which is formulated as follows:

% \begin{equation}
% soft^{\bm{\tau}^k}(\bm{x}) = sgn(\bm{x})(|\bm{x}| - \bm{\tau}^k).
% \end{equation}

\noindent\textbf{Regional Dynamic Aggregation.}
%\JN{Before getting the output of multiple blocks, we need to assemble different transformations to different regions.}
We propose a regional dynamic aggregation strategy to aggregate the fundamental transformation into a dynamic mixed domain through region-based feature scoring. 

% The regional dynamic aggregation is used to assemble the different transformations to different regions through the region-based feature scoring, and then aggregated into a dynamic mixed domain for reconstruction.

%Given the output from the multiple blocks, next is to 

\paragraph{Regional Weighting}
We begin with extracting regional spatial information through local average pooling:
\begin{equation}
    \bm{P}_i^k = Pool_s(\bm{R}_i^k),
\end{equation}
where $s$ denotes the pooling kernel size.
As shown in Fig.~\ref{Structure}, we take $\bm{R}^k$ as input to retain low-level details for scoring and finalize the regional feature extraction with average pooling.
%\JN{As shown in Fig.~\ref{Structure}, we take $\bm{R}^k$ as the input to retain low-level details and provide information for scoring, using average pooling to finish the regional extraction.}

Intuitively, a larger pooling kernel will introduce more average information and thus lose regional characteristics that determine the transformation domain. 
While a small pooling kernel may introduce redundancy and increased computational  overhead. We set the pooling  kernel size as $s=5$, as discussed in Sec.~\ref{ablation} c).

%\paragraph{ScoreNet}
Next, we establish a mapping from region characteristics to transformation domains. To this end, following the proposed regional dynamic FISTA algorithm, the regional dynamic weight $w_i^k$ is computed as:
\begin{equation}
\bm{w}_i^k = Softmax(\bm{\omega}_2(\sigma(\bm{\omega}_1 \cdot \bm{P}_i^k + \bm{b}_1)) + \bm{b}_2).\\
\end{equation}
We design a ScoreNet to learn coefficients $\left\{\bm{w}_i^k\right\}_i^N$ to static FISTA transformation domains, which helps to produce dynamic sparse representations fitting to different regions. Specifically, we use two convolutional layers with a activation layer to discriminate different regions' features and apply a softmax activation to generate normalized attention weights $\bm{w}_i^k$ for each dynamic block.

%which is detailed as follows:
\begin{table*}[]
    \centering
    \caption{PSNR(dB) comparison of the test methods on 10 scenes in the simulation dataset.}
\begin{tabular}{l|llllllllll|l}
\hline
{{Method}}                & {{Scene1}}         & {{Scene2}}         & {{Scene3}}         & {{Scene4}}         & {{Scene5}}         & {{Scene6}}         & {{Scene7}}         & {{Scene8}}         & {{Scene9}}         & {{Scene10}}        & {{Average}}        \\ \hline
{{TwIST~\cite{TwIST}}}                 & {{25.16}}          & {{23.02}}          & {{21.40}}          & {{30.19}}          & {{21.41}}          & {{20.95}}          & {{22.20}}          & {{21.82}}          & {{22.42}}          & {{22.67}}          & {{21.12}}          \\
{{GAP-TV~\cite{GAP-TV}}}                & {{26.82}}          & {{22.89}}          & {{26.31}}          & {{30.65}}          & {{23.64}}          & {{21.85}}          & {{23.76}}          & {{21.98}}          & {{22.63}}          & {{23.10}}          & {{24.36}}          \\
{ADMM-TV~\cite{ADMM-TV}}               & {25.77}          & {21.39}          & {23.14}          & {33.70}          & {23.43}          & {23.68}          & {18.62}          & {23.39}          & {23.25}          & {23.86}          & {24.02}          \\
{PNP-HSI~\cite{PNP-CASSI}}               & {26.35}          & {22.60}          & {26.78}          & {37.61}          & {24.88}          & {24.85}          & {20.12}          & {23.80}          & {25.11}          & {24.57}          & {25.67}          \\
{{DeSCI~\cite{DeSCI}}}                 & {{27.15}}          & {{22.26}}          & {{26.56}}          & {{39.00}}          & {{24.80}}          & {{23.55}}          & {{20.03}}          & {{20.29}}          & {{23.98}}          & {{25.94}}          & {{25.86}}          \\
{DeepRED~\cite{Deepred}}               & {28.27}          & {21.64}          & {24.42}          & {37.93}          & {25.04}          & {26.14}          & {22.62}          & {23.42}          & {28.35}          & {25.62}          & {26.35}          \\
{U-Net~\cite{U-Net}}                 & {28.28}          & {24.06}          & {26.02}          & {36.33}          & {25.51}          & {27.97}          & {21.15}          & {26.83}          & {26.13}          & {25.07}          & {26.80}          \\
{{HSSP~\cite{HyperNet}}}                  & {{31.07}}          & {{26.30}}          & {{29.00}}          & {{38.24}}          & {{27.98}}          & {{29.16}}          & {{24.11}}          & {{27.94}}          & {{29.14}}          & {{26.44}}          & {{28.93}}          \\
{{$\lambda$-Net~\cite{lambda-net}}}                 & {{30.82}}          & {{26.30}}          & {{29.42}}          & {{37.37}}          & {{27.84}}          & {{30.69}}          & {{24.20}}          & {{28.86}}          & {{29.32}}          & {{27.66}}          & {{29.25}}          \\
{{TSA-Net~\cite{TSA-Net}}}               & {{31.26}}          & {{26.88}}          & {{30.03}}          & {{39.90}}          & {{28.89}}          & {{31.30}}          & {{25.16}}          & {{29.69}}          & {{30.03}}          & {{28.32}}          & {{30.15}}          \\
{{DNU~\cite{DNU}}}                   & {{31.72}}          & {{31.13}}          & {{29.99}}          & {{35.34}}          & {{29.03}}          & {{30.87}}          & {{28.99}}          & {{30.13}}          & {{31.03}}          & {{29.14}}          & {{30.74}}          \\
{DIP-HSI~\cite{DIP-HSI}}               & {32.68}          & {27.26}          & {31.30}          & {40.54}          & {29.79}          & {30.39}          & {28.18}          & {29.44}          & {34.51}          & {28.51}          & {31.26}          \\
{GAP-Net~\cite{GAP-Net}}               & {33.03}          & {29.52}          & {33.04}          & {\textbf{41.59}} & {30.95}          & {32.88}          & {27.60}          & {30.17}          & {32.74}          & {29.73}          & {32.13}          \\
{{GSM~\cite{GSM}}}                   & {{33.26}}          & {{32.09}}          & {{33.06}}          & {{40.54}}          & {{28.86}}          & {{33.08}}          & {{\textbf{30.74}}} & {{31.55}}          & {{34.66}}          & {{31.44}}          & {{32.63}}          \\ \hline
{{RDFNet(Ours)}} & {{\textbf{33.40}}} & {{\textbf{32.38}}} & {{\textbf{34.47}}} & {{37.70}}          & {{\textbf{32.67}}} & {{\textbf{35.80}}} & {{27.67}}          & {{\textbf{33.09}}} & {{\textbf{34.66}}} & {{\textbf{31.54}}} & {{\textbf{33.34}}} \\ \hline
\end{tabular}
\label{PSNR}
\end{table*}
\begin{table*}[]
    \centering
    \caption{SSIM comparison of the test methods on 10 scenes in the simulation dataset.}
\begin{tabular}{l|llllllllll|l}
\hline
{Method}       & {Scene1}         & {Scene2}         & {Scene3}         & {Scene4}         & {Scene5}         & {Scene6}         & {Scene7}         & {Scene8}         & {Scene9}         & {Scene10}        & {Average}        \\ \hline
{TwIST~\cite{TwIST}}        & {0.700}          & {0.604}          & {0.711}          & {0.851}          & {0.635}          & {0.644}          & {0.643}          & {0.650}          & {0.690}          & {0.569}          & {0.669}          \\
{GAP-TV~\cite{GAP-TV}}       & {0.754}          & {0.610}          & {0.802}          & {0.852}          & {0.703}          & {0.663}          & {0.688}          & {0.654}          & {0.682}          & {0.584}          & {0.699}          \\
{ADMM-TV~\cite{ADMM-TV}}      & {0.729}          & {0.589}          & {0.737}          & {0.834}          & {0.699}          & {0.648}          & {0.603}          & {0.631}          & {0.682}          & {0.559}          & {0.671}          \\
{PNP-HSI~\cite{PNP-CASSI}}      & {0.712}          & {0.613}          & {0.786}          & {0.877}          & {0.721}          & {0.685}          & {0.648}          & {0.691}          & {0.687}          & {0.611}          & {0.703}          \\
{DeSCI~\cite{DeSCI}}        & {0.794}          & {0.694}          & {0.877}          & {0.965}          & {0.778}          & {0.753}          & {0.772}          & {0.740}          & {0.818}          & {0.666}          & {0.785}          \\
{DeepRED~\cite{Deepred}}      & {0.769}          & {0.602}          & {0.769}          & {0.927}          & {0.757}          & {0.743}          & {0.777}          & {0.674}          & {0.840}          & {0.721}          & {0.758}          \\
{U-Net~\cite{U-Net}}        & {0.822}          & {0.777}          & {0.857}          & {0.877}          & {0.795}          & {0.794}          & {0.799}          & {0.796}          & {0.804}          & {0.710}          & {0.803}          \\
{HSSP~\cite{HyperNet}}         & {0.852}          & {0.798}          & {0.875}          & {0.926}          & {0.827}          & {0.823}          & {0.851}          & {0.831}          & {0.822}          & {0.740}          & {0.834}          \\
{$\lambda$-Net~\cite{lambda-net}}        & {0.880}          & {0.846}          & {0.916}          & {0.962}          & {0.866}          & {0.886}          & {0.875}          & {0.880}          & {0.902}          & {0.843}          & {0.886}          \\
{TSA-Net~\cite{TSA-Net}}      & {0.887}          & {0.855}          & {0.921}          & {0.964}          & {0.878}          & {0.895}          & {0.887}          & {0.887}          & {0.903}          & {0.848}          & {0.893}          \\
{DNU~\cite{DNU}}          & {0.863}          & {0.846}          & {0.845}          & {0.908}          & {0.833}          & {0.887}          & {0.839}          & {0.885}          & {0.876}          & {0.849}          & {0.863}          \\
{DIP-HSI~\cite{DIP-HSI}}      & {0.890}          & {0.833}          & {0.914}          & {0.962}          & {0.900}          & {0.877}          & {0.913}          & {0.874}          & {0.927}          & {0.851}          & {0.894}          \\
{GAP-Net~\cite{GAP-Net}}      & {0.921}          & {0.903}          & {0.940}          & {0.972}          & {0.924}          & {0.927}          & {0.921}          & {0.904}          & {0.927}          & {0.901}          & {0.924}          \\
{GSM~\cite{GSM}}          & {0.915}          & {0.898}          & {0.925}          & {0.964}          & {0.882}          & {0.937}          & {0.886}          & {0.923}          & {0.911}          & {0.925}          & {0.917}          \\ \hline
{RDFNet(Ours)} & {\textbf{0.950}} & {\textbf{0.954}} & {\textbf{0.961}} & {\textbf{0.976}} & {\textbf{0.957}} & {\textbf{0.963}} & {\textbf{0.939}} & {\textbf{0.956}} & {\textbf{0.958}} & {\textbf{0.949}} & {\textbf{0.956}} \\ \hline
\end{tabular}
\label{SSIM}
\end{table*}

\paragraph{Dynamic Aggregation.}
We obtain the dynamic sparse representation of RDFNet by softly assembling the output of multiple dynamic blocks $\bm{x}_i^k$ based on the region-based coefficients $\bm{w}_i^k$ predicted by ScoreNet.
\begin{equation}
    \bm{X}^k = \sum_i^N(\bm{X}_i^k \cdot \bm{w}_i^k).\\
\end{equation}
As a result, RDFNet constructs the sparse transformation in a dynamic data-driven manner for different regions. The core weight coefficients $\bm{W}^k = \sum_i^N\bm{w}_i^k$ are learned adaptively according to region's characteristic.
The regional adaptive transformation enables our dynamic blocks with more flexibility in reconstruction compared to previous works. 
% \JN{Our dynamic blocks gain more flexibility than others in terms of reconstructions with the region-based transformation adjusting strategy.}
% \JN{Our dynamic blocks gain flexibility in reconstructions with the region-based transformation adjusting strategy.} 

\subsection{Learning Objectives}
Given the training data pairs ${(\bm{y}_j,(\bm{x}_{gt})_j)^D_{j=1}}$, RDFNet takes the measurement $\bm{y}$ as input and generates the reconstruction $\bm{x}$. We seek to reduce the discrepancy between $\bm{x}$ and $(\bm{x}_{gt})$, which indicates the accuracy of the inverse function, while satisfying the symmetry constraint in each dynamic block. Furthermore, we measure the sparsity of spectral frames in the learned domain.
% Assume there are $K$ phases in total, $\bm{x}^k$ for $k\in{K}$ is the output of the $k$-th phase, (such description shoud have been mentioned in earlier in the paragrpah we detail the overall framework of the model. K is not determined or introduced by the loss function, can not introduce or assume it here)
For the output $\bm{X}^k$ in the $k$-th phase, denote $\bm{X}_{gt}$ as the tensor form of the groundtruth $\bm{x}_{gt}$, we design the loss function for RDFNet as:\\
\begin{equation}
L_{acc} = ||\bm{X} - \bm{X}_{gt}||^2_2,
\end{equation}
\begin{equation}
L_{sym} = \frac{1}{K}\sum_{k=1}^K\sum_{i=1}^N||{F'_i}(F_i(\bm{X}_i^k)) - \bm{X}_i^k ||^2_2, 
\end{equation}
\begin{equation}
L_{spa} = \frac{1}{K}\sum_{k=1}^K||\hat F(\bm{X}^k)||_1.
\end{equation}
% \JN{The total loss function consists of the above \R{three} terms in proportion to each other:}
The final loss is a weighted combination of the above three terms:
% \JN{The total loss function consists of the above three terms in proportion to each other, summed up with weight:}
\begin{equation}
    L_{all} = \alpha \cdot L_{acc} + \beta \cdot L_{spa} + \gamma \cdot L_{sym},
\end{equation}
where $\alpha$, $\beta$ and $\gamma$ are balancing coefficients. By default, we set $\alpha = 1$, $\beta = 0.01$ and $\gamma = 0.001$.
%$N$ is the number of dynamic FISTA block, here $N=3$.
\begin{table*}
		\caption{{Model size, computation, performance and speed comparison on the KAIST~\cite{KAIST} dataset. The size of the test input spectral cube is $256\times256\times28$. All the other settings are kept the same for a fair comparison. Best results are in bold.}}
		\small
		\begin{center}
			\setlength{\tabcolsep}{2mm}
			\begin{tabular}{l|ccccc|c}
			\toprule[1.0pt]
				    &{$\lambda$-Net~\cite{lambda-net}}&{TSA-Net~\cite{TSA-Net}}&{DNU~\cite{DNU}}&{DIP-HSI~\cite{DIP-HSI}}&{GSM~\cite{GSM}}&{RDFNet(Ours)} \\
                    \midrule[{0.75pt}]
                    {Params(M)}&{62.64}&{44.25}&{4.63}&{33.85}&{3.76}&{\textbf{1.29}}\\
                     \midrule[{0.75pt}]
                    {FLOPs(G)}&{117.98}&{110.06}&{606.32}&{\textbf{64.42}}&{646.35}&{604.88}\\
                    \midrule[{0.75pt}]    
                    {PSNR(dB)}&{28.53}&{31.46}&{30.74}&{31.26}&{32.63}&{\textbf{33.34}}\\
                    \midrule[{0.75pt}]
                    {SSIM}&{0.841}&{0.894}&{0.863}&{0.894}&{0.917}&{\textbf{0.956}}\\
                    \midrule[{0.75pt}]
                    {Time(s)}&{0.13}&{4.07}&{2.74}&{4.95}&{0.22}&{\textbf{0.11}}\\
            \bottomrule[1pt]
			\end{tabular}
		\end{center}
		\label{Param}
\end{table*}

\begin{figure*}[]
  \centering   
  \includegraphics[width=1.0\linewidth]{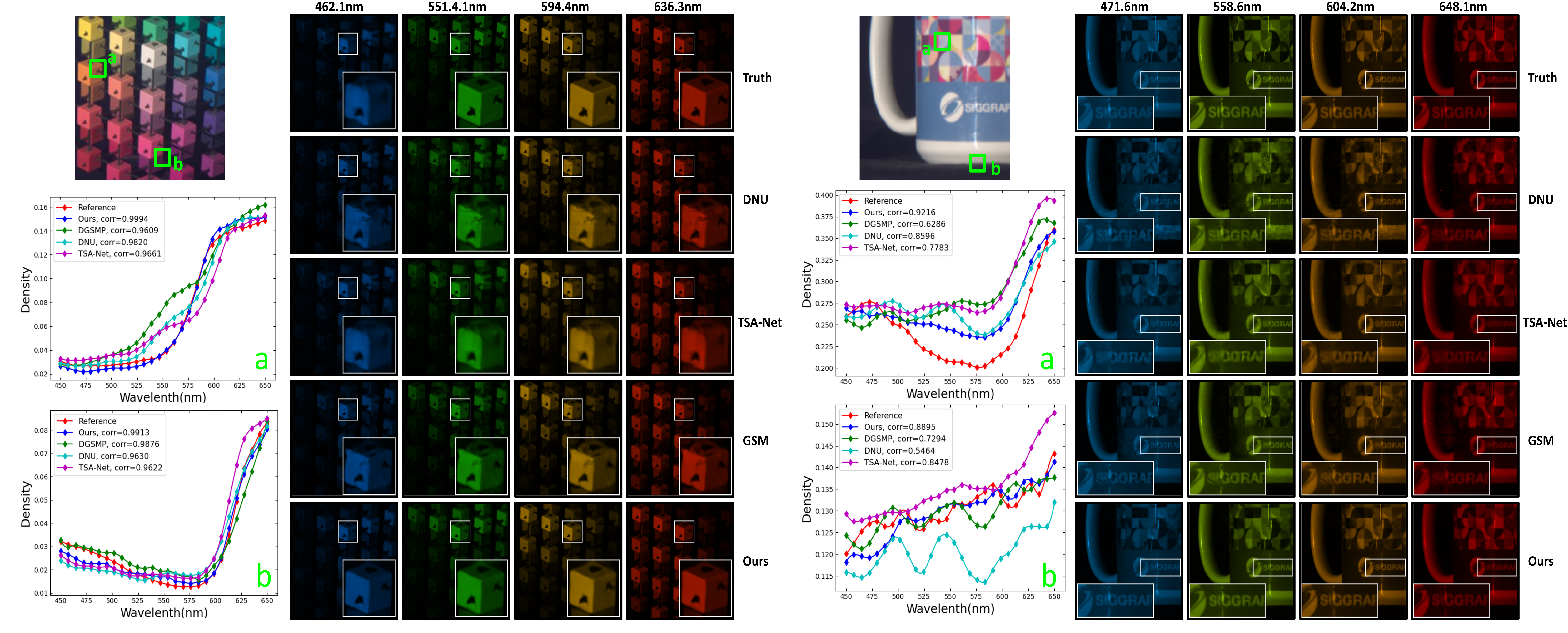}
    \caption{Reconstructed images of scene2 and scene5 with 4 out of 28 spectral channels by the three deep learning-based methods. A region in each scene are selected for analysing the spectra of the reconstructed results. Zoom in for better view.}
   \label{fig:simu_results}
\end{figure*}

\section{Experiment}
% This section introduces our experimental settings, including the implementation details and the used evaluation metrics. Then, 
We evaluate our RDFNet on both simulated and real data and report the evaluation of parameters, FLOPs, and inference speed. Extensive ablation studies are further provided to validate our design choices and parameter settings. 

\subsection{Experimental Settings}
We unfold the proposed iterative algorithm into five phases. Each phase contains one RDFNet. All experiments are conducted on a NVIDIA RTX-3090. 
We set the number of dynamic block as 3 and the regional pooling kernel size as $5\times5$.
We train the model for $3,000$ epochs using  Adam optimizer~\cite{adam}  with learning rate 0.0001 and batch size 4.
The Peak-Signal-to-Noise Ratio (PSNR) and structural similarity index (SSIM)~\cite{metrics} are employed to evaluate the quality of reconstructed spectral data-cube.

\subsection{Results on Simulated Data}
\paragraph{Data and setups}
% \subsection{Datasets and Evaluation Metrics}
We conduct simulations on three popular hyperspectral image datasets including CAVE~\cite{CAVE}, KAIST~\cite{KAIST} and ICVL~\cite{ICVL}. 
For CAVE~\cite{CAVE} and KAIST~\cite{KAIST}, similar to TSA-Net~\cite{TSA-Net} and GSM~\cite{GSM}, we employ the real mask of size $256\times256$ for simulation. Following TSA-Net~\cite{TSA-Net} and GSM~\cite{GSM}, we train the model on CAVE and test on $10$ $256\times256$ sized scenes extracted from KAIST. 
To keep consistent with the wavelengths in real systems~\cite{TSA-Net}, we unify the wavelength of train and test data by spectral interpolation. Thus, the modified train and test data have $28$ spectral bands ranging from $450\rm{nm}$ to $650\rm{nm}$.

The ICVL~\cite{ICVL} dataset consists of $201$ real-world objects, each with $1300\times1392$ spatial resolution and $31$ spectral bands collected from $400\rm{nm}$ to $700\rm{nm}$ in a $10\rm{nm}$ step.
For ICVL~\cite{ICVL}, we follow the procedure in HSCNN~\cite{HSCNN} and DNU~\cite{DNU}. Similar to KAIST~\cite{KAIST} and CAVE~\cite{CAVE}, we select $28$ spectral bands ranging from $450\rm{nm}$ to $650\rm{nm}$ for training and testing. We set the image size as $1024\times1024$ for training and randomly collect $10$ $256\times256$ sized images from ICVL for testing.

\begin{table*}
		\caption{{Comparison results of the proposed network and state-of-the-art HSI reconstruction methods on the ICVL dataset. Best results are in bold.}}
		\small
		\begin{center}
			\setlength{\tabcolsep}{2mm}
			\begin{tabular}{l|ccccccc|c}
			\toprule[1.0pt]
				&{{TwIST~\cite{TwIST}}}&{{TV~\cite{tv}}}&{{$\lambda$-Net~\cite{lambda-net}}}&{{HSCNN~\cite{HSCNN}}}&{{ISTA~\cite{ISTA-Net}}}&{Low-rank~\cite{R1-1}}&{{DNU~\cite{DNU}}}&{{RDFNet(Ours)}} \\
			\midrule[{0.75pt}]
				{PSNR(dB)}&{26.15}&{25.44}&{29.01}&{28.45}&{30.50}&{30.92}&{32.61}&{\textbf{35.51}}\\
				\midrule[{0.75pt}]
				{SSIM}&{0.936}&{0.906}&{0.946}&{0.934}&{0.947}&{0.874}&{\textbf{0.966}}&{0.961}\\
            \bottomrule[1pt]
			\end{tabular}
		\end{center}
		\label{ICVL}
\end{table*}

\paragraph{Comparisons with SOTAs.}
We compare our proposed Regional Dynamic FISTA-Net with several state-of-the-art HSI reconstruction algorithms on the dataset KAIST~\cite{KAIST}, including three traditional methods (TwIST~\cite{TwIST}, GAP-TV~\cite{GAP-TV}, and ADMM-TV~\cite{ADMM-TV}), two model based methods (PNP-HSI~\cite{PNP-CASSI} and DeSCI~\cite{DeSCI}), three prior based methods (DeepRED~\cite{Deepred}, HSSP~\cite{HyperNet}, and DIP-HSI~\cite{DIP-HSI}) and six deep learning based methods (U-Net~\cite{U-Net}, $\lambda$-net~\cite{lambda-net}, TSA-Net~\cite{TSA-Net}, DNU~\cite{DNU}, GAP-Net~\cite{GAP-Net}, and GSM~\cite{GSM}).

% traditional method; TwIST, GAP-TV, ADMM-TV
% prior based method; DeepRED, HSSP, DIP-HSI
% model based; PNP-HSI, DeSCI,  
% deep learning based; U-Net, lambda-Net, TSA-Net, DNU, GAP-Net, GSM

% All the performance for competitive methods is released publicly. 
% We cite all experimental results from GSM~\cite{GSM} and all comparison methods are trained on the same training dataset. 
The PSNR and SSIM results of different methods on 10 scenes in the simulation datasets are listed in Tab.\uppercase\expandafter{\romannumeral1} and Tab.\uppercase\expandafter{\romannumeral2}. The params FLOPs, and inference time of open-source CNN-based algorithms are reported in Tab.\uppercase\expandafter{\romannumeral3}. It can be observed from these three tables that our RDFNets significantly surpass previous methods by a large margin on all 10 scenes while requiring much cheaper memory and computational costs. More specifically, our RDFNet surpasses the leading algorithm GSM~\cite{GSM}, DIP-HSI~\cite{DIP-HSI}, DNU~\cite{DNU} and TSA-Net~\cite{TSA-Net} by 0.71, 2.08, 2.6, and 3.19 dB, and 0.039, 0.062, 0.093, and 0.063 SSIM, while costing 34.3\% (1.29/3.76), 3.8\%, 27.9\% and 2.9\% Params and 50.0\% (0.11/0.22), 2.2\%, 4.0\% and 2.7\% inference time.

In particular, RDFNet achieves promising performance with only less than $35\%$ parameters compared to the second-best GSM~\cite{GSM}. Meanwhile, the inference time of RDFNet is only $0.11$ second per image, demonstrating clear superiority over prior state-of-the-arts in terms of both accuracy and efficiency.

Since our method is based on deep unfolding and requires multiple phases of calculation, it has more FLOPs than the end-to-end TSA-Net~\cite{TSA-Net}, $\lambda$-Net~\cite{lambda-net} or the prior based methods DIP-HSI~\cite{DIP-HSI}. While it has the least FLOPs compared to other deep unfolding algorithms including DNU~\cite{DNU} and GSM~\cite{GSM}.
% \R{Tab.\uppercase\expandafter{\romannumeral1} and Tab.\uppercase\expandafter{\romannumeral2} list the reconstruction results on KAIST.} 
% Our method outperforms other model based and deep based priors by a large margin.
% \R{Specifically, it surpasses the second-best GSM~\cite{GSM} by $0.71$ dB in average PSNR and $0.0396$ in average SSIM.}
% Compared to other deep unfolding methods, our method improves TSA-Net~\cite{TSA-Net} and DNU~\cite{DNU} by $1.88$ and $2.60$ dB in average PSNR and $0.0623$ and $0.0931$ in average SSIM, respectively.

Fig.~\ref{fig:simu_results} demonstrates the details and spectral curves of the reconstructed HSIs. The recovered spectral images are converted to synthetic-RGB (sRGB) via the CIE color matching function. It can be seen that our method have more edge details and less undesirable visual artifacts than those from other methods. And the reconstructed spectral curves of the proposed methods have a higher correlation with the reference spectra. Moreover, one can see from Fig.~\ref{fig:simu_results} that satisfactory shape reconstruction results have been achieved at the edge of the cube, and the text outlines on the cup body are well reconstructed with their depths close to reality.
%\JN{Besides, it can be seen that in the enlarged reconstruction details from Fig.~\ref{fig:simu_results}, satisfactory shape reconstruction result has been achieved for the edge of the box, and the outlines of the text on the cup body are well reconstructed with their depths close to reality.}
% \R{We compare the spatial details and spectral accuracy of the above 8 algorithms on Scene 9, with results shown in Fig.4. The recovered spectral images are converted to synthetic-RGB (sRGB) via the CIE color matching function. It can be seen that the optimization algorithms suffer from the blurry on the horizontal axis, which might be caused by the shifting effects of the disperser in the system. PnP-HSI is unable to fully exert its advantage due to the less-than-perfect initialization of ADMM-TV. Compared with DeepRED and TSA-Net, the results of our PnP-DIP show sharper edges and better visual qualities. In addition, the reconstructed spectral curves of the proposed methods have a higher correlation with the reference spectra.}

Surprisingly, on the other simulation datasets ICVL, Our method outperforms all the priors. The results are listed in Tab.\uppercase\expandafter{\romannumeral4}.
% The best results for each index are highlighted in bold. 

Specifically, compared to model based methods, the proposed regional dynamic network better captures the distinct characteristic of HSI. Our method also produces a remarkable improvement upon learning based priors. The boost upon RDFNet evidences that the regional dynamic transformation with adaptive thresholds is more conducive for HSI reconstruction than the fixed transformation with manually-set thresholds. Noticeably, our method outperforms other methods by $9.36\rm{dB}$ (TwIST~\cite{TwIST}), $10.07\rm{dB}$ (TV~\cite{tv}), $6.5\rm{dB}$ ($\lambda$-Net~\cite{lambda-net}), $7.06\rm{dB}$ (HSCNN~\cite{HSCNN}), $5.01\rm{dB}$ (ISTA~\cite{ISTA-Net}), $4.59\rm{dB}$ (Low-rank~\cite{R1-1}) and $2.9\rm{dB}$ (DNU~\cite{DNU}) in average PSNR.

% \JN{In summary, the small model sizes and low inference time mean that the model is very fast to reconstruction and can still be explored to a greater extent using higher computing power of the GPU.}

% \R{In a nutshell, from the above visualization results and quantitative indicators, we can see that our proposed method outperforms all the other reconstruction methods, demonstrating its effectivenes and advantages.}

\begin{figure*}[htbp]
  \centering
   \includegraphics[width=1.0\linewidth]{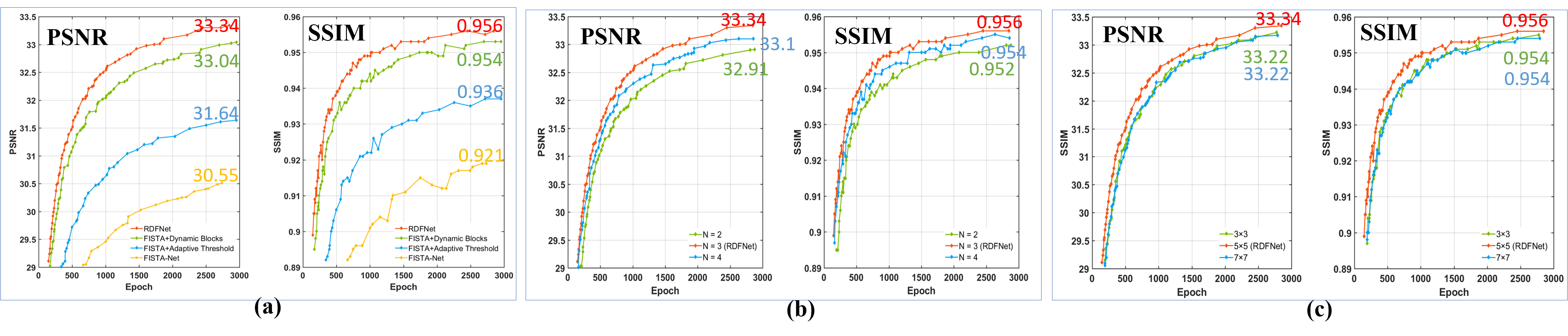}
   \caption{Ablation results. Ablations of (a) regional dynamic block and pixel-wise adaptive soft-thresholding; (b) number of the regional dynamic blocks; (c) regional pooling size. }
   \label{ablationstudy}
\end{figure*}

\subsection{Results on Real Data}

\begin{figure}[]
  \centering
    \includegraphics[width=0.9\linewidth]{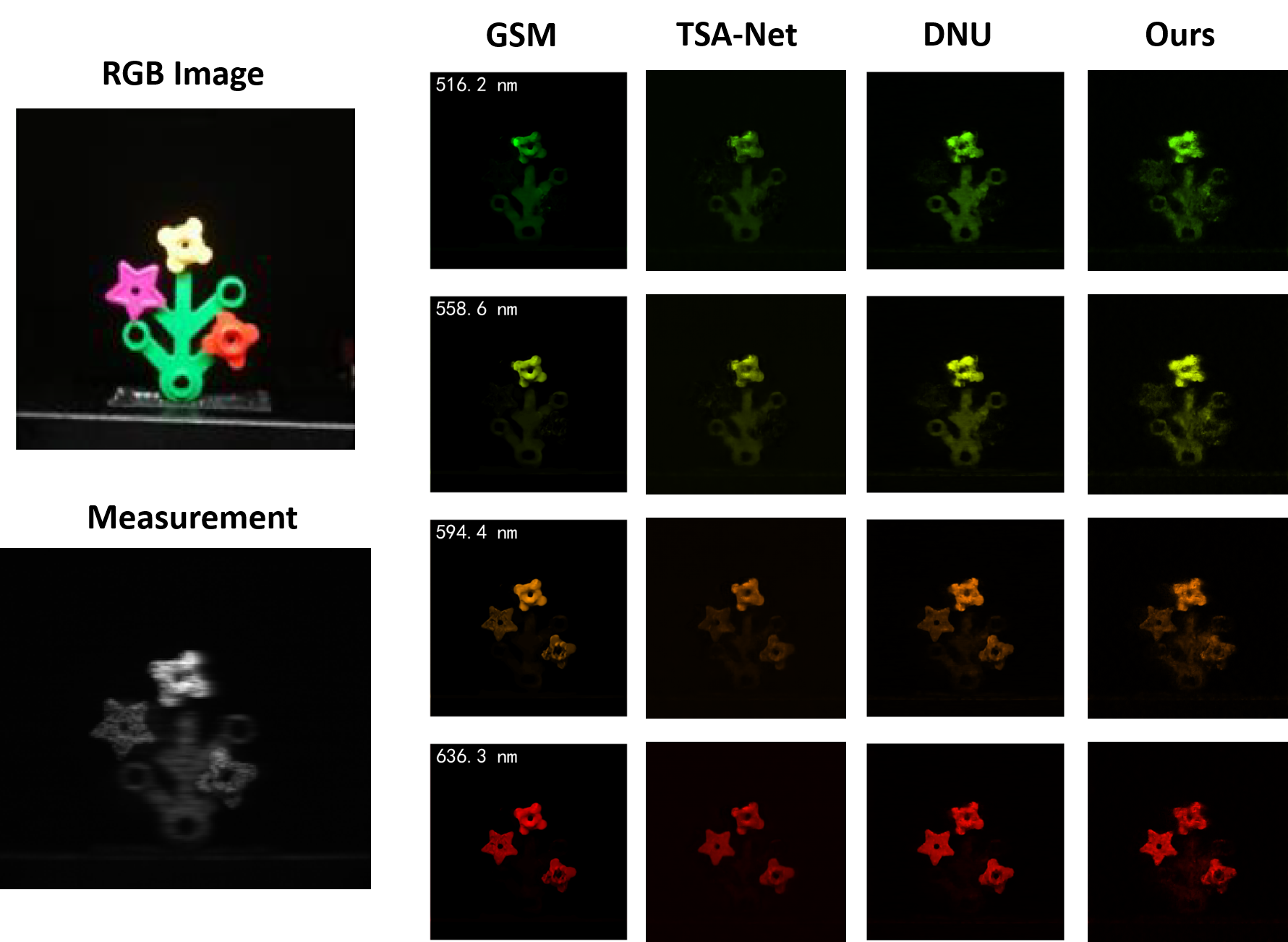}
   \caption{Reconstructed images of the real scene1 with 4 our of 28 spectral channels by the competing methods.}
   \label{real}
\end{figure}

% \paragraph{Datasets and Settings.}
% \subsection{Real Data Results}
We test our methods on real SD-CASSI data~\cite{burst, TSA-Net} that captures real scenes with $28$ wavelengths ranging from $450\rm{nm}$ to $650\rm{nm}$ and has 54-pixel dispersion in the column dimension. Thus, the measurements captured by the system have a spatial size of $660\times714$.
% \paragraph{Comparisons with SOTAs.}
Fig.~\ref{real} shows the reconstruction results of scene1 with four channels by RDFNet and other competing methods. One can observe that our method well recovers textures in both spectral and spatial dimensions.

%\JN{give more analysis. }

\begin{figure}[htbp]
  \centering
\includegraphics[width=0.8\linewidth]{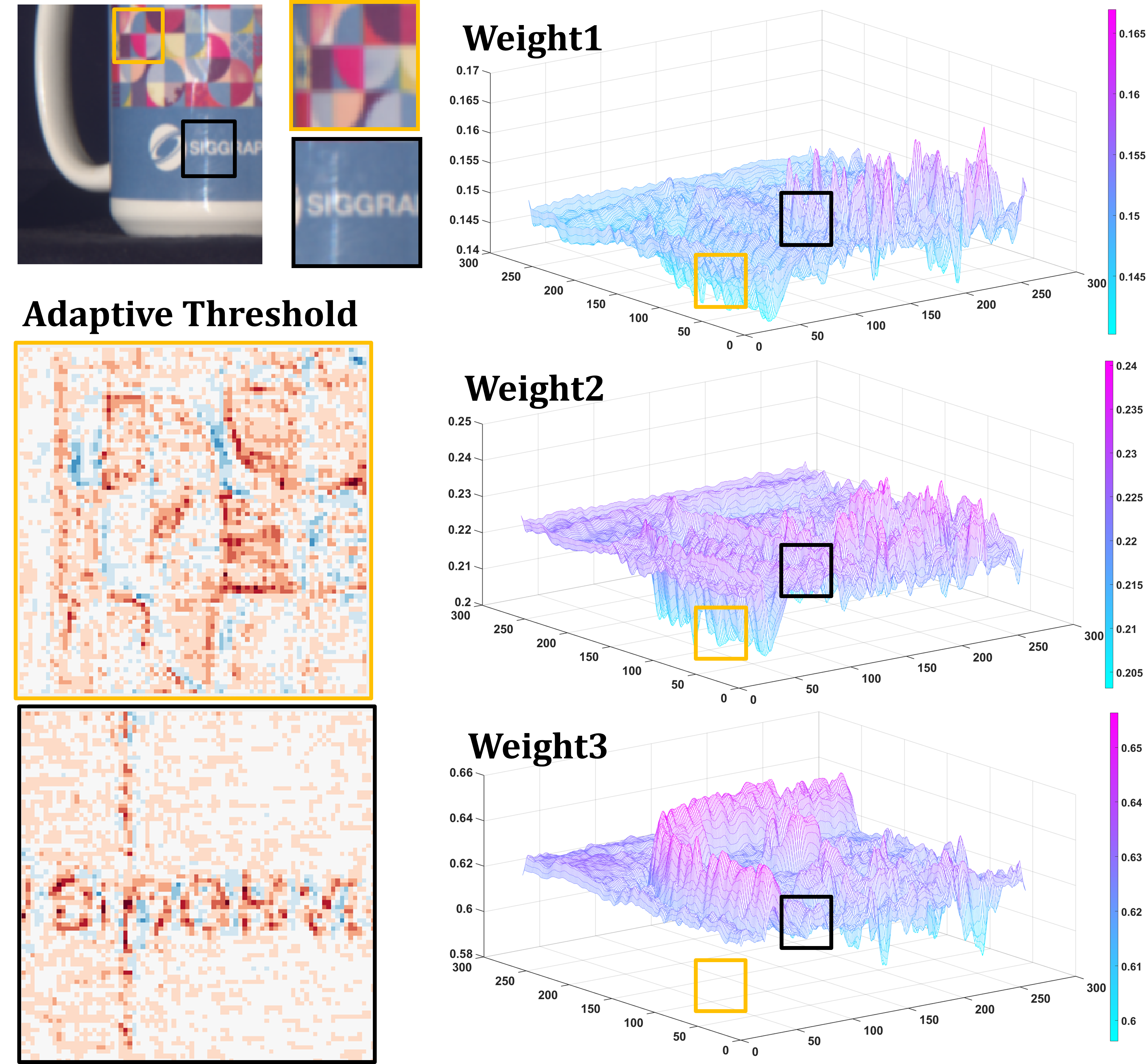}
   \caption{Visualization of dynamic weights in each dynamic block and adaptive soft-thresholding in RDFNet. We randomly selected two regions that were labeled respectively.}
   \label{fig:visualization}
\end{figure}

\subsection{Ablation Studies}
\label{ablation}
% Fig.~\ref{ablationstudy} shows the ablation results, \R{the analysis is as follows}:

% \subsubsection{Ablations on components}
%\noindent\textbf{Effect of key components.}
\noindent\textbf{Effect of key components.}
The regional dynamic FISTA network consists of two key components: the region-based dynamic block used to transform different patches into different sparse domains and the pixel-wise adaptive thresholding module used to dynamically determinate appropriate shrinkage scale. 
% In order to verify whether the two modules contribute to the network, we designed a component ablation experiment to test. 
We test the effectiveness of each of the two components by incorporating them one-by-one progressively.
%We add modules dynamic FISTA block and adaptive soft-thre
% shold to the baseline in turn, and then add them to baseline togeter.

As shown in Fig.~\ref{ablationstudy} (a), the quality of reconstruction (evaluated by PSNR and SSIM) is gradually increasing. Our RDFNet achieves the best performance and outperforms the FISTA-Net baseline by $2.79\rm{dB}$ in average PSNR. Tab.~\ref{tabablation} shows the improvements by separately incorporating the dynamic block and adaptive soft-thresholding are $0.3\rm{dB}$ and $1.7\rm{dB}$ in average, respectively. It indicates that the regional dynamic strategy largely contributes to the performance gain and the adaptive soft-thresholding brings additional improvement. 
% The ablation experiment results are almost consistent with our expectations. Both the proposed two components are essential for the network's performance.

		\begin{table}
		\arrayrulecolor{black}
		\caption{Ablation analysis of key components. \textbf{AT} represents Adaptive Threshold, \textbf{DB} represents Dynamic Blocks.}
		\small
		\begin{center}
			\setlength{\tabcolsep}{2mm}
			\begin{tabular}{l|cc|ccc}
			\toprule[1.0pt]
				&AT&DB&PSNR&SSIM \\
			\midrule[0.75pt]
				Baseline~\cite{FISTA-Net}&&&30.55&0.921\\
				Baseline+AT&\checkmark&&31.64&0.936\\
				Baseline+DB&&\checkmark&33.04&0.954\\
				\midrule[0.75pt]
				RDFNet(Ours)&\checkmark&\checkmark&\textbf{33.34}&\textbf{0.956}\\
            \bottomrule[1pt]
			\end{tabular}
		\end{center}
		\label{tabablation}
	\end{table}	

\noindent\textbf{Impact of block number.}
\label{block}
% Previous work shows that the number of dynamic components is related to the network's performance~\cite{Dynamic}. 
To investigate the impact of dynamic block number $N$, we test the model variants with $N=2, 3, 4$. 
Fig.~\ref{ablationstudy} (b) shows that our model is not sensitive to the number of dynamic blocks affects reconstruction error but only to a certain extent. 
The model with $N=3$ achieves the best in both PSNR and SSIM. Decreasing the block number, \textit{i.e.}, $N=2$, leads to slight performance degradation. One possible explanation is that fewer blocks result in fewer transformation domains for dynamic regulation. In addition, increasing the blocks \textit{i.e.}, $N=4$, brings no further performance improvement. The reason is that too many parameters may cause the problem of poor network convergence. 
% Nevertheless, our methods show good robustness in changing the parameters.

\noindent\textbf{Impact of region pooling.}
\label{pool}
We further study the impact of the kernel size of regional pooling in Fig.~\ref{ablationstudy} (c).
We test the model variants with pooling kernel size of $3\times3$, $5\times5$, and $7\times7$. 
The model with $5\times5$ pooling performs the best. A smaller $3\times3$ pooling kernel, which leads to more fine-grained region division, affects the extraction of regional characteristics and thus hinders the dynamic adjustment of transformation domain. While a larger $7\times7$ kernel blurs the region division and leads to unreasonable weights allocation.

\section{Conclusion}
We have proposed a regional dynamic FISTA algorithm for coded aperture snapshot spectral imaging. Unlike the existing static transformation network, we develop a novel hierarchical regional dynamic structure that adjusts different regions into adaptive transformations according to their characteristics. Besides, a pixel-wise attention strategy have been used on soft-thresholding. Extensive experiments show that the proposed RDFNet achieves the best reconstruction results, demonstrating clear superiority over prior state-of-the-arts in terms of both accuracy and efficiency. Specifically, the proposed RDFNet achieves an average PSNR of $35.51\rm{dB}$ among seven mainstream methods on the ICVL~\cite{ICVL} and $33.34\rm{dB}$ among fourteen kinds of HSI reconstruction methods on the KAIST~\cite{KAIST}. While on the parameters analysis, our proposd method achieves only 1.29M parameters and inference time of 0.11 second per-image and obtains competitive results on the FLOPs.
% Furthermore, our proposed method RDFNet achieves only 1.29M parameters and inference time of 0.11 second per image, which demonstrates clear superiority over prior state-of-the-arts in terms of both accuracy and efficiency.}

Our proposed method is not limited to spectral SCI. It can also be used in video SCI systems. One future direction of interest is to extend the dynamic transform domain to other tasks. 
% The other direction is to search the ways of using our proposed method for the actual application.

\section*{Acknowledgments}
% This work was supported in part by the National Key Scientific Instrument and Equipment Development Project of China (61527802). And Jianan Li is supported by Beijing Institute of Technology Research Fund Program for Young Scholars.

This work was financially supported by the National Key Scientific Instrument and Equipment Development Project of China (No. 61527802), the National Natural Science Foundation of China (No. 62101032),  the Postdoctoral Science Foundation of China (Nos. 2021M690015, 2022T150050), and Beijing Institute of Technology Research Fund Program for Young Scholars (No. 3040011182111).

\bibliographystyle{IEEEtran}
\bibliography{egbib.bib}

\begin{IEEEbiography}[{\includegraphics[width=1in,height=1.25in,clip,keepaspectratio]{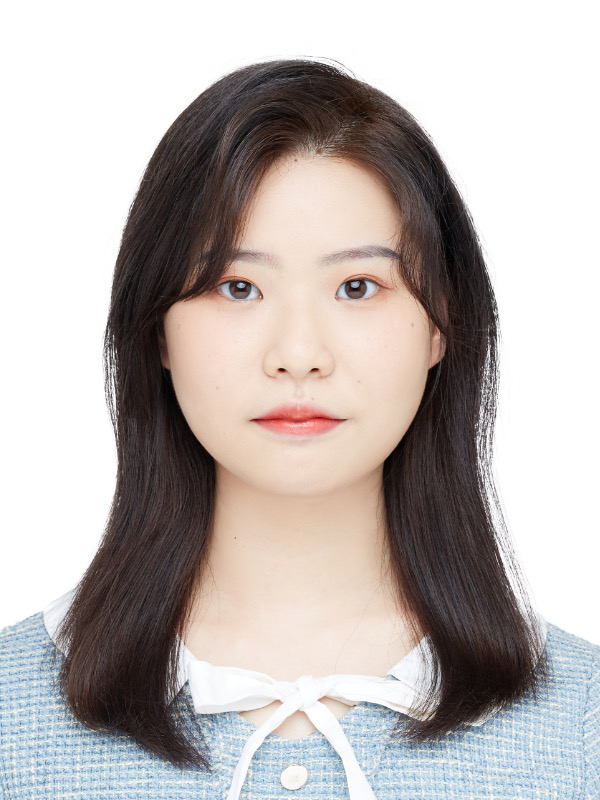}}]{Shiyun Zhou} is currently pursuing the M.S. degree with the School of Optics and Photonics, Beijing Institute of Technology, Beijing, China. Her research interests include hyperspectral image processing and compressive spectral reconstruction.
\end{IEEEbiography}

\begin{IEEEbiography}[{\includegraphics[width=1in,height=1.25in,clip,keepaspectratio]{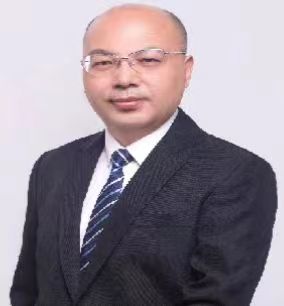}}]{Tingfa Xu} received the Ph.D. degree from the Changchun Institute of Optics, Fine Mechanics and Physics, Changchun, China, in 2004. He is currently a Professor with the School of Optics and Photonics, Beijing Institute of Technology, Beijing, China. His research interests include optoelectronic imaging and detection and hyperspectral remote sensing image processing.
\end{IEEEbiography}

\begin{IEEEbiography}[{\includegraphics[width=1in,height=1.25in,clip,keepaspectratio]{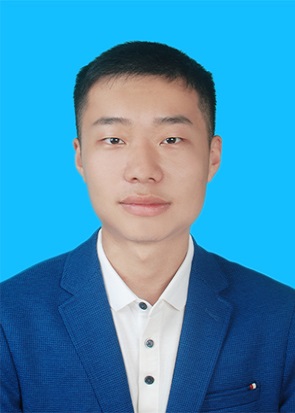}}]{Shaocong Dong} is currently pursuing the M.S. degree with the School of Optics and Photonics, Beijing Institute of Technology, Beijing, China. His research interests include hyperspectral image processing, 3d point cloud data processing and compressive spectral reconstruction.
\end{IEEEbiography}

\begin{IEEEbiography}[{\includegraphics[width=1in,height=1.25in,clip,keepaspectratio]{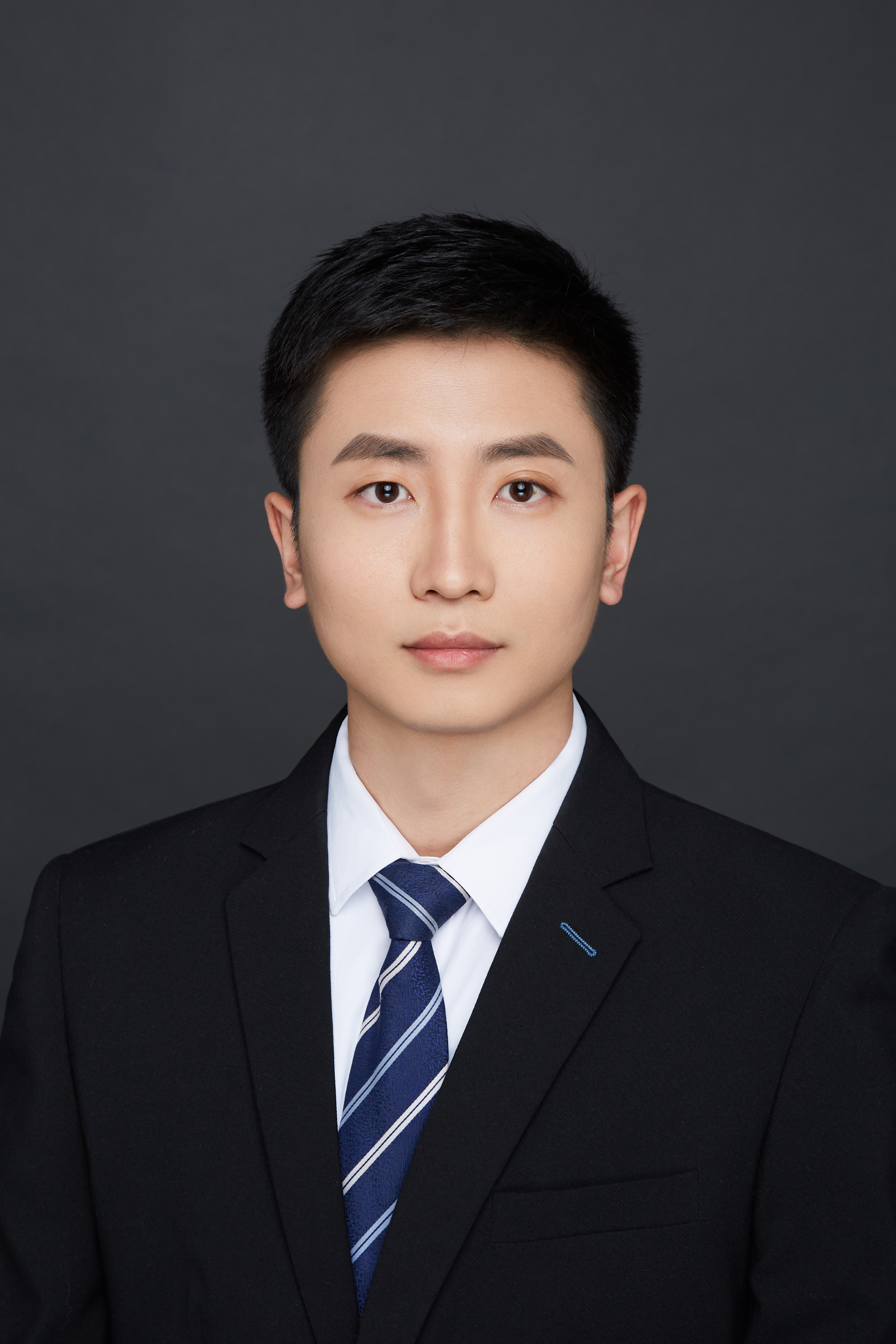}}]{Jianan Li} is currently an assistant professor at School of Optics and Photonics, Beijing Institute of Technology, Beijing, China, where he received his B.S. and Ph.D. degree in 2013 and 2019, respectively. From July 2015 to July 2017, he worked as a joint training Ph.D. student at National University of Singapore. From October 2017 to April 2018, he worked as an intern at Adobe Research. His research interests mainly include computer vision and real-time image/video processing.
\end{IEEEbiography}

\vfill

\end{document}